\documentclass[8pt]{revtex4}
\pdfoutput=1
\usepackage{amssymb}
\usepackage{latexsym}
\usepackage{epsfig}
\begin{document}

\title{ Emergence of $F(R)$ gravity-analogue due to defects in graphene}

\author{Alireza Sepehri $^{1,2}$}
\email{alireza.sepehri@uk.ac.ir} \affiliation{ $^{1}$Faculty of
Physics, Shahid Bahonar University, P.O. Box 76175, Kerman,
Iran.\\$^{2}$ Research Institute for Astronomy and Astrophysics of
 Maragha (RIAAM), P.O. Box 55134-441, Maragha, Iran. }

\author{Richard Pincak$^{3,4}$}\email{pincak@saske.sk}
\affiliation{ $^{3}$ Institute of Experimental Physics, Slovak Academy of Sciences,
Watsonova 47,043 53 Kosice, Slovak Republic}
\affiliation{ $^{4}$ Bogoliubov Laboratory of Theoretical Physics, Joint
Institute for Nuclear Research, 141980 Dubna, Moscow region, Russia}

 \author{Ahmed Farag Ali}
\email{ahmed.faragali@nias.knaw.nl} \affiliation{Deptartment of Physics, Faculty of Science, Benha University, Benha 13518, Egypt.}
\affiliation{Netherlands Institute for Advanced Study, Korte Spinhuissteeg 3, 1012 CG Amsterdam, ​
Netherlands.}

\begin{abstract}
We show that the defects of graphene, which lead to the non-equality between positive curvature of fermions with anti-parallel spins and negative curvature of fermions with parallel spins, imply an emergence of $F(R)$ gravity. By increasing the number of atoms at each defect, the order of scalar curvature increases and the shape of $F(R)$ gravity  changes. This gravity has a direct relation with  energy-momentum tensor and leads to motion of electrons in a special path and hence producing superconductivity. Also, for some special angles, parallel spins become close to each other and repel to each other. In that condition, the shape of  $F(R)$ gravity changes and electrons can't continue to move in an initial path and return. Consequently, superconductivity disappears  and one new conductivity appears in opposite direction. \\\\


 \end{abstract}
 \date{\today}


 \maketitle

\section{Introduction}
Recently, graphene has attracted many attentions because  it has shown the ability to make a transition from the semiconductivity  to the superconductivity state \cite{s2,s3,q1,q2,q5}. For example, some investigations have shown that
graphene with (effective) vacancy disorder is a physical representative of dirty d-wave superconductors. Also, the density of states (DoS) of this system has been obtained numerically and within the self-consistent T-matrix approximation (SCTMA) in the presence of vacancies \cite{s2}. Besides, some studies investigated the quantum size effects in armchair graphene nano-ribbons (AGNR) with hydrogen termination using density functional theory (DFT) in Kohn-Sham formulation. They have obtained the electronic structure of this system  and predicted a threefold periodicity of the excitation gap with ribbon width \cite{s3}. In another paper, it has been argued that the best way for creating the superconductor in temperature of Room is a constructing of a wormhole inside the graphene \cite{q1}.  A graphene wormhole is composed of  a short nanotube which behaves like a bridge between two different graphene layers. Other papers have discussed that graphene can be applied as a platform to demonstrate a  wormhole� for electrons. They have shown that two properly designed graphene-based nanomaterials can effectively annihilate� one another from an electronic point of view, and  similar to a wormhole, lead to the delocalization of the wave-function stationary states \cite{q2}. And in more recent works,  it has been shown that the wormhole in graphene can be produced by joining photons that exchange between atoms and acts similar to nano-BIon in M-theory. This BIonic wormhole is formed by decreasing the separation distance between layers of graphene and enhancing the cooper hopping pairing between them. Also, it has been argued that external magnetic field produces a new BIon which causes losing the energy density of initial-BIon between two layers and disappearing the superconductivity \cite{q5}.

Now, the question arises how curvature of graphene changes during its evolution from semiconductor to superconductor? We will show that during this transition, higher order of curvature scalars are produced and $F(R)$ gravity emerges. In cosmology, the shape of $F(R)$ gravity,  which is the main cause of evolution of universe, has been determined \cite{m1,m2,m6,v1,v2,v3,v4,v5,v6}. For example, It was argued
that for producing the bounce cosmology, the $F(R)$ gravity should
have the function with positive rational numbers powers of the Ricci scalar
($R$) in the large cosmic time regime and a Gauss hyper geometric function
in the small cosmic time \cite{m1}. Some other authors found that the $F(R)$
gravity which creates the  super-bounce has the form $R + \alpha R^{2}$ at the
large- curvature regime and $R + c_{1}R^{-1/2}+\Lambda$ at the small-curvature
regime \cite{m2}. Also,
it has been discussed that a Type IV singular bounce can be produced by the $F(R)$ gravity
which has the form $F(R)= R+\alpha R^{2}$  near the singularity\cite{m6}.  Some other authors have proposed a mechanism that allows to reconcile dark energy models with $F(R)$ theories and determines  the shape of $F(R)$ which leads to the same Hubble parameter of the given quintessence model and astrophysical data \cite{v1}.  In another paper, authors have obtained the relations between the model independent results and parameters coming from cosmography and the theoretically motivated assumptions of $F(R)$ cosmology \cite{v2}.  In another scenario, authors considered the electrically charged black holes in the Palatine formalism of  $F(R)$ gravity. They have shown that $F(R)= R\pm\alpha R^{2}/R_{P}$ where $R_{P}$ is the planck curvature may lead to different geometrical structures  in terms of inner horizon for black holes \cite{v5}. In another investigation, some authors have shown that Palatini  $F(R)$ gravity yields a geometry in which point like singularity of space-time in four dimensional classical scenarios is replaced by a finite size wormhole structure. In fact, this theory causes to the emergence of nonsingular space-time, despite the existence of curvature divergences at the wormhole throat \cite{v6}.  In another research, authors discussed that microscopic wormholes which are produced in modified gravity theories like $F(R)$ gravity play a role analogous to that of defects in crystals. This illustrates how lessons from solid state physics can be used in modified theories of gravity \cite{v3}. Finally, authors studied the metric-affine scenarios where an action in modified gravity theories like $F(R)$ gravity  is generated by electrovacuum fields in a three dimensional space-time. Such models are can be applied in crystalline structures with microscopic defects and, in particular, those that may be effectively behaved as bi-dimensional (like graphene) \cite{v4}. Extending these results to the graphene, we propose a new model which considers the process of formation of an special $F(R)$ gravity as due to defects  which its shape depends on the shape of defects and the number of atoms in graphene.  In usual graphene, each carbon has four electrons which bound to  three electrons of other atoms and one electron is free. This electron may move along the sheet of graphene and produces the conductivity. However, in this system, the curvature produced by non-identical fermions is neutralized by the curvature of identical fermions and total curvature of system is zero. The curvature has a direct relation with energy momentum and thus sum over applied momentum on electron is zero. This means that  free electrons don't move in a certain path and no have any effect on conductivity. By entering defects, for some angles between sides, anti-parallel spins become close to each other
 and curvature produced by their couplings will be more respect to curvature produced by parallel spins. In these conditions, one momentum is applied to free electrons and leads to their motion in one special path and consequently, the superconductivity emerges. For some other angles, parallel spins become close to each other and repel each other. For these defects, a new $F(R)$ gravity emerges that shape of it is different from previous state. Also, the couplings of anti-parallel spins in curvatures are removed and the sign of curvatures reverses which is a signature of anti-gravity in this system. \\
 In this paper, we will use of M-theory language to describe complex systems such as graphene. Maybe the question arises that what is  the advantage of using that approach? In the description of graphene and its properties, one just needs standard fields such as fermions, scalars, and gauge bosons. To answer this question, we should say that only in M-theory, we can consider the origin of Pauli exclusion principle which has the main role in conducting of free electrons and show that the reason for attractive force between anti-parallel spins is the appearance of gravity between them and the repelling force between parallel spins is due to the emergence of anti-gravity between them. In fact, in M-theory, at the first, there are only scalar fields that are attached to M0-branes. Then, by joining M0-branes to each other and the formation of M1-branes, some scalar fields which are attached to them transit to gauge fields. Also, by breaking the symmetry of M1-branes such as the upper and lower parts of M1-branes are not the same, gauge fields are broken and fermions are produced. Thus, the origin of all three types of matter like scalars, fermions and gauge fields are the same and we can write all interactions in terms of one field only. In a complicated system like graphene, this method makes the consideration and calculations easier respect to the usual field theory that there are three types of indipendent fields. Also, in this method, each free electron interacts with other electrons which some of them are parallel and some are anti-parallel respect to it. If the curvatures produced by parallel spins are neutralized by the curvatures produced by anti-parallel spins, the applied force to electrons is zero, free electrons move randomly and conductivity decreases. However, if curvatures of anti-parallel spins be more than the curvatures of parallel spins, electrons move in one special direction and conductivity increases.\\
The outline of the paper is the  following.  In section \ref{o1}, we discuss how,
by entering defects, the curvature of non-parallel spins increases and $F(R)$ gravity emerges.   In section \ref{o2}, we argue that by changing properties of graphene, the curvature of parallel spins becomes more than curvature of anti-parallel spin and anti-$F(R)$ gravity emerges. This leads to the transmission of electrons in opposite direction respect to expected path. The last section is devoted to discussion and conclusions.

\section{Emergence of $F(R)$ gravity as due to defects
}\label{o1}
In this section, we design a model which considers the interaction between electrons in a graphene and calculate
the curvature produced with different particles in this system. To this end, first we assume that all electrons are paired in the shape of scalars. Then, we break this symmetry and produce gauge fields and in third stage, we break symmetry again and produce fermions. We will show that the curvature produced by parallel spins is not neutralized by the curvature ($R_{\mu\nu}$) produced by non-parallel spins and $F(R)$ gravity emerges. With respect to this point that energy-momentum tensors ($T_{\mu}^{\nu}=diag[-p,-p,-p,\rho]$) have direct relations with curvatures ($R_{\mu\nu}-\frac{1}{2}R g_{\mu\nu}=T_{\mu\nu}$), the sum over applied momentum on the electron  is not zero and consequently, electrons  move in one special path.

In a graphene, three electrons of each atom is paired with three electrons of other atom and make spineless pairs. These pairs can be known as scalars in this system and will be shown by "X" in the action. Free electrons are represented by $\psi$ and electrons in each pair are displayed by $\Psi$. Also, gauge fields which are exchanged between electrons are the main cause of gravity between anti-parallel spins and anti-gravity between parallel spins and shown by $A^{\mu\nu}$. Molecules in each graphene are composed of six atoms which are located on the corners of one hexagonal system. We assume that at the beginning, the graphene is symmetric and all the electrons are paired and at rest and no any gauge field is exchanging between them. There are many similarities between a graphene and branes. For this reason, we use of the model in \cite{q6,q7} for M-theory, to write the relevant action for symmetric state of graphene. In cosmology, we have shown that at the beginning, in the world, we only have M0-branes which are zero-dimensional objects that scalars are attached to them. Only, after joining these M0-branes to each other and formation of M1-brane, gauge fields and fermions are emerged and symmetry is broken. Then, usual action of Mp-branes are produced which include all three types of scalars, gauge fields and fermions \cite{qq8}. In this paper, we proposed a model which scalars are pairs of electrons and there is a very high symmetry at first. Then symmetry is broken and fermions and gauge fields are produced and shape of this action changes to usual action.  Using the method in \cite{q6,q7,qq8,q8}, we introduce the following action for conducting from each atom in graphene \cite{q6,q7}:

\begin{eqnarray}
S_{co-Atom} = \int dt Tr( \Sigma_{M,N,L=0}^{p}
\langle[X^{M},X^{N},X^{L}],[X^{M},X^{N},X^{L}]\rangle) \label{s1}
\end{eqnarray}

where $X^{M}=X^{M}_{\alpha}T^{\alpha}$, T's are generators of 3-Li algebra \cite {q6,q7,q8,qq8,q9,q10,q11,q12}  and

\begin{eqnarray}
 &&[T^{\alpha}, T^{\beta}, T^{\gamma}]= f^{\alpha \beta \gamma}_{\eta}T^{\eta} \nonumber \\&&\langle T^{\alpha}, T^{\beta} \rangle = h^{\alpha\beta} \nonumber \\&& [X^{M},X^{N},X^{L}]=[X^{M}_{\alpha}T^{\alpha},X^{N}_{\beta}T^{\beta},X^{L}_{\gamma}T^{\gamma}]\nonumber \\&&\langle X^{M},X^{M}\rangle = X^{M}_{\alpha}X^{M}_{\beta}\langle T^{\alpha}, T^{\beta} \rangle
\label{s2}
\end{eqnarray}

Here,  $X^{M}$(M=1,2,...p) denote the pairs in graphene, $h^{\alpha\beta}$ is the metric of system and p is the number of pairs around each carbon. To obtain total action of conductivity in a graphene, we should sum over actions of all atoms and  use of following action:

\begin{eqnarray}
&&S_{co-Graphene} = \int  d^{3}x \sum_{n=1}^{Q}\beta_{n}\Big(
\delta^{a_{1},a_{2}...a_{n}}_{b_{1}b_{2}....b_{n}}L^{b_{1}}_{a_{1}}...L^{b_{n}}_{a_{n}}\Big)^{1/2}.\nonumber\\&&
(L)^{a}_{b}=\delta_{a}^{b} Tr(  \Sigma_{a,b=0}^{p}\Sigma_{j=p+1}^{Q\times (p-1)}\Big(
\langle[X^{a},X^{i},X^{j}],[X_{a},X_{i},X_{j}]\rangle+
\langle[X^{a},X^{c},X^{d}],[X_{b},X_{c},X_{d}]\rangle+\nonumber\\&&\langle[X^{a},X^{b},X^{j}],[X_{b},X_{a},X_{j}]\rangle+\langle[X^{k},X^{i},X^{j}],[X_{k},X_{i},X_{j}]\rangle
\Big)) \label{s3}
\end{eqnarray}

 where a,b are indices of pairs on each atom, i,j are indices of free pairs related to free electrons, Q is the number of atoms and p is the number of pairs in each atom.  Now, we break the initial symmetry for the first time and  assume that pairs move around the atoms and interact with each other via gauge fields. In fact, these gauge fields are produced as due to break of initial symmetry. To this end, we use of following relations \cite {q6,q7,q8,qq8,q9,q10,q11,q12}:

 \begin{eqnarray}
 &&\langle[X^{a},X^{i},X^{j}],[X^{a},X^{i},X^{j}]\rangle=
 \frac{1}{2}\varepsilon^{abc}\varepsilon^{abd}(\partial_{c}X^{i}_{\alpha})(\partial_{d}X^{i}_{\beta})
 \langle(T^{\alpha},T^{\beta})\rangle \sum_{j} (X^{j})^{2} =
  \frac{1}{2}\langle \partial^{a}X^{i},\partial_{a}X^{i}\rangle \sum_{j} (X^{j})^{2}=\nonumber \\&&
    \frac{1}{2}\langle \partial^{a}X^{i},\partial_{a}X^{i}\rangle \sum_{j} F(X^{j})\nonumber \\
 && F(X)=\sum_{j} (X^{j})^{2}\nonumber \\
  &&\nonumber \\
 &&\nonumber \\
 &&\langle[X^{a},X^{b},X^{j}],[X^{a},X^{b},X^{j}]\rangle=
  \frac{1}{2}\varepsilon^{abc}\varepsilon^{ade}(\partial_{b}\partial_{c}X^{i}_{\alpha})(\partial_{e}\partial_{d}X^{i}_{\beta})
  =  \frac{1}{2}\langle \partial^{b}\partial^{a}X^{i},\partial_{b}\partial_{a}X^{i}\rangle\nonumber \\
   &&\nonumber \\
  &&\nonumber \\
  &&\langle[X^{a},X^{b},X^{c}],[X^{a},X^{b},X^{c}]\rangle=
 -\lambda^{2}
 (F^{abc}_{\alpha\beta\gamma})(F^{abc}_{\alpha\beta\eta})\left(\langle[T^{\alpha},T^{\beta},T^{\gamma}],[T^{\alpha},T^{\beta},T^{\eta}]
 \rangle\right)=\nonumber \\
 &&-\lambda^{2}
 (F^{abc}_{\alpha\beta\gamma})(F^{abc}_{\alpha\beta\eta})f^{\alpha
 \beta \gamma}_{\sigma}h^{\sigma \kappa}f^{\alpha \beta
 \eta}_{\kappa} \langle T^{\gamma},T^{\eta}\rangle =-\lambda^{2}
 (F^{abc}_{\alpha\beta\gamma})(F^{abc}_{\alpha\beta\eta})\delta^{\kappa
 \sigma} \langle T^{\gamma},T^{\eta}\rangle= -\lambda^{2} \langle
 F^{abc},F^{abc}\rangle \nonumber\\
 \nonumber\\&& F_{abc}=\partial_{a} A_{bc}-\partial_{b}
 A_{ca}+\partial_{c}
 A_{ab}. \nonumber \\
 &&\nonumber \\
 && i,j=p+1,..,p\times Q\quad a,b=0,1,...p\quad m,n=0,..,P\times Q~~ \label{s4}
 \end{eqnarray}

 In above relation, $A_{ab}$ is $2$-form gauge field which exchange between pairs
 and $\lambda^{2}$ is related to the angle between electrons respect to center of molecule in each graphene and it's dimension is of order $\frac{1}{l_{a}^{3}}$ where $l_{a}$ is the relative distance between two atoms.  Also, indices on X's are very important. When indices of X's in each bracket be a,b, c which are indices of bound electrons, brackets produce different results respect to brackets which include indices i,j,k. Substituting mappings of equation (\ref{s4}) in equation (\ref{s3}), we obtain:

 \begin{eqnarray}
 &&S_{co-Graphene} = \int  d^{3}x \sum_{n=1}^{Q}\beta_{n}\Big(
 \delta^{a_{1},a_{2}...a_{n}}_{b_{1}b_{2}....b_{n}}L^{b_{1}}_{a_{1}}...L^{b_{n}}_{a_{n}}\Big)^{1/2}.\nonumber\\&&
 (L)^{a}_{b}=\delta_{a}^{b} Tr(  \Sigma_{a,b=0}^{p}\Sigma_{j=p+1}^{Q\times (p-1)}\Big(
    \frac{1}{2}\langle \partial_{a}X^{i},\partial_{a}X^{i}\rangle F (X^{j})+ \nonumber\\&&\frac{1}{2}\langle \partial^{b}\partial^{a}X^{i},\partial_{b}\partial_{a}X^{i}\rangle-\lambda^{2} \langle
     F^{abc},F_{abc}\rangle +\langle[X^{k},X^{i},X^{j}],[X_{k},X_{i},X_{j}]\rangle
 \Big)) \label{s5}
 \end{eqnarray}

It is clear from above equation that by regarding the interaction between electrons, the symmetry of system is broken and gauge fields  are produced. In fact by entering gauge fields, the shape of action changes to usual actions in M-theory. This action is an approximation of following action \cite{q6,q7,qq8,q8,q9,q10,q11,q12,q13,q14,q15,q16,q17,q18}:

  \begin{eqnarray}
  &&S_{co-Graphene} = \int  d^{3}x \sum_{n=1}^{Q}\beta_{n}\Big(
  \delta^{a_{1},a_{2}...a_{n}}_{b_{1}b_{2}....b_{n}}L^{b_{1}}_{a_{1}}...L^{b_{n}}_{a_{n}}\Big)^{1/2}\nonumber\\&&
  (L)^{a}_{b}=\delta_{a}^{b}STr (-det(P_{abc}[ E_{mnl}
  +E_{mij}(Q^{-1}-\delta)^{ijk}E_{kln}]+ \lambda
  F_{abc})det(Q^{i}_{j,k}))\label{ss5}
  \end{eqnarray}

 where

 \begin{eqnarray}
  E_{mnl}^{\alpha,\beta,\gamma} &=& G_{mnl}^{\alpha,\beta,\gamma} + B_{mnl}^{\alpha,\beta,\gamma}, \nonumber\\
  Q^{i}_{j,k} &=& \delta^{i}_{j,k} + i\lambda[X^{j}_{\alpha}T^{\alpha},X^{k}_{\beta}T^{\beta},X^{k'}_{\gamma}T^{\gamma}]
  E_{k'jl}^{\alpha,\beta,\gamma},\nonumber\\F_{abc}&=&\partial_{a} A_{bc}-\partial_{b} A_{ca}+\partial_{c}
 A_{ab}. \label{ss4}
 \end{eqnarray}
 and $X^{M}=X^{M}_{\alpha}T^{\alpha}$, $A_{ab}$ is $2$-form gauge
 field,

 \begin{eqnarray}
  &&[T^{\alpha}, T^{\beta}, T^{\gamma}]=
  f^{\alpha \beta \gamma}_{\eta}T^{\eta} \nonumber \\&& [X^{M},X^{N},X^{L}]=
  [X^{M}_{\alpha}T^{\alpha},X^{N}_{\beta}T^{\beta},X^{L}_{\gamma}T^{\gamma}]
 \label{sss5}
 \end{eqnarray}

 where $\lambda=2\pi l_{s}^{2}$,
 $G_{mnl}=g_{mn}\delta^{n'}_{n,l}+\partial_{m}X^{i}\partial_{n'}X^{i}\sum_{j}
 (X^{j})^{2}\delta^{n'}_{n,l} + \frac{1}{2}\langle \partial^{b}\partial^{a}X^{i},\partial_{b}\partial_{a}X^{i}\rangle $ and $X^{i}$ refers to scalar field.

Now, we break the symmetry of initial system again and consider the interaction between electrons in each pair. Each scalar is constructed from two fermions with upper and lower spins ($\psi$ or $\Psi$) and each gauge field ($F^{abc}$) is exchanged between two fermions. We can define \cite{qq8}:

\begin{eqnarray}
&& A_{ab}\rightarrow \psi^{U}_{a}\psi^{L}_{b}-\psi^{L}_{a}\psi^{U}_{b}\nonumber \\
&&\partial_{a} =\partial_{a}^{U}+\partial_{a}^{L}\nonumber \\
&&\partial_{a}^{U}\psi^{U}_{a}=1,\;\partial_{a}^{L}\psi^{L}_{a}=1\label{st6}
\end{eqnarray}

These definitions help us to break the symmetry and produce fermions. In this model, each gauge field can be written in terms of fermions which interact with it such as the antisymmetric state has been regarded. Previously in  \cite{qq8}, it has been shown that by using this definition, the Pauli exclusion principle has been re-obtained and Dirac equations for Fermions have been derived.  Using this technique, we obtain \cite{qq8}:

\begin{eqnarray}
&&\langle
F^{abc},F_{abc}\rangle_{Free-Free}=A^{ab}\partial_{a}^{U}\psi^{L}_{b}-A^{ab}\partial_{a}^{L}\psi^{U}_{b}+\nonumber \\
&&-\psi^{\dag a, L}\psi^{L}_{a}-\psi^{\dag a, U}\psi^{U}_{a}+\psi^{\dag a, L}\psi^{U}_{a}+\psi^{\dag a, U}\psi^{L}_{a}-\nonumber \\
&&(\psi^{\dag a, U}\partial^{a,U}\psi^{L}_{a})(\psi^{\dag a, U}\partial_{a}^{U}\psi^{L}_{a})-(\psi^{\dag a, L}\partial_{a}^{L}\psi^{a,U})(\psi^{\dag a, L}\partial_{a}^{L}\psi^{U}_{a})-\nonumber \\
&&(\psi^{\dag a, U}\partial^{a,U}\psi^{L}_{a})(\psi^{\dag a, L}\partial_{a}^{L}\psi^{U}_{a})-(\psi^{\dag a, L}\partial^{a,L}\psi^{U}_{a})(\psi^{\dag a, U}\partial_{a}^{U}\psi^{L}_{a})+\nonumber \\
                  &&[(\psi^{\dag a, U}\partial^{a,L}\psi^{U}_{a})(\psi^{\dag a, U}\partial_{a}^{L}\psi^{U}_{a})+
                  (\psi^{\dag a, L}\partial^{a,U}\psi^{L}_{a})(\psi^{\dag a, L}\partial_{a}^{U}\psi^{L}_{a})]=\nonumber \\&&A^{ab}i\sigma_{ij}^{2}\partial_{a}^{i}\psi^{j}_{b}+\sigma^{0}_{ij}\psi^{\dag a,i}\psi^{j}_{a}-\sigma^{1}_{ij}\psi^{\dag a, i}\psi^{j}_{a}+\nonumber \\
&&\sigma^{0}_{i'i}(\psi^{\dag a, i'}i\sigma^{0}_{i'j}\sigma^{1}_{jk}\partial^{a,j}\psi^{k}_{a})(\psi^{\dag  i}_{a}i\sigma^{0}_{ij}\sigma^{1}_{jk}\partial^{a,j}\psi^{k}_{a})-\nonumber \\
&&\sigma^{1}_{i'i}(\psi^{\dag a, i'}i\sigma^{0}_{i'j}\sigma^{1}_{jk}\partial^{a,j}\psi^{k}_{a})(\psi^{\dag  i}_{a}i\sigma^{0}_{ij}\sigma^{1}_{jk}\partial^{a,j}\psi^{k}_{a})-\nonumber \\
&&\sigma^{0}_{i'i}(\psi^{\dag a, i'}i\sigma^{1}_{i'j}\sigma^{1}_{jk}\partial^{a,j}\psi^{k}_{a})(\psi^{\dag  i}_{a}i\sigma^{1}_{ij}\sigma^{1}_{jk}\partial^{a,j}\psi^{k}_{a})\nonumber \\\nonumber \\\nonumber \\
&&\langle
F^{abc},F_{abc}\rangle_{tot}=\langle
F^{abc},F_{abc}\rangle_{Free-Free}+\langle
F^{abc},F_{abc}\rangle_{Free-Bound}+\langle
F^{abc},F_{abc}\rangle_{Bound-Bound}\label{s6}
\end{eqnarray}

where we have used of definition of $\sigma_{ij}^{1}=\Big{(}\begin{array}{cc}
0 & 1 \\
1 & 0
\end{array}\Big{)} $, $\sigma_{ij}^{0}=\Big{(}\begin{array}{cc}
1 & 0 \\
0 & 1
\end{array}\Big{)} $, $\sigma_{ij}^{2}=\Big{(}\begin{array}{cc}
 & -i \\
i & 0
\end{array}\Big{)} $,  and i,j=U,L. refers to upper and lower spins.  Above equation shows that by breaking symmetry of graphene, usual equations of Dirac fields are produced and the  Pauli matrices has been created. In fact, this method helps us to obtain the exact form of interaction between fermions without adding any thing by hand. Another interesting thing that comes out from this equation is the relation between gauge fields and their sources which are fermions. These results are very similar to those in M-theory \cite{q6,q7,qq8}. When, M0-branes join to each other completely, only gauge fields are emerged, however if they link non-completely, the symmetry of system is broken. In these conditions, the upper and lower sections of M1-brane that is formed by joining M0-branes are different and fields that are attached to upper section produces the fermions with upper spin and those which are linked to lower section produces the fermion with lower spin \cite{qq8}. In a graphene, the existence of defect causes that the symmetry is broken. In these conditions, number of upper or lower spins in some points of graphene increases and coupling of anti-parallel spins is more than coupling of parallel spins. Consequently, the curvature of anti-parallel spins are not neutralized by curvature of parallel spins and gravity in this graphene emerges. In this model, each scalar is composed of two anti-parallel spins which are connected by a gauge field ($X\rightarrow \psi^{U}_{a}A^{ab}\psi^{L}_{b}-\psi^{L}_{a}A_{ab}\psi^{U}_{b}$). Using this definition, we can calculate other terms in action (\ref{ss5})\cite{qq8}:

\begin{eqnarray}
&& X\rightarrow \psi^{U}_{a}A^{ab}\psi^{L}_{b}-\psi^{L}_{a}A^{ab}\psi^{U}_{b}+\Psi^{U}_{a}A^{ab}\Psi^{L}_{b}-\Psi^{L}_{a}A^{ab}\Psi^{U}_{b}+\Psi^{U}_{a}A^{ab}\psi^{L}_{b}-\psi^{L}_{a}A^{ab}\Psi^{U}_{b}\nonumber \\
&&\nonumber \\
&&\partial_{a} =\partial_{a}^{U}+\partial_{a}^{L}\nonumber \\
&&\partial_{a}^{U}\psi^{U}_{a}=1,\;\partial_{a}^{L}\psi^{L}_{a}=1\nonumber \\
&&\nonumber \\
&&\langle \partial^{b}\partial^{a}X^{i},\partial_{b}\partial_{a}X^{i}\rangle=\varepsilon^{abc}\varepsilon^{ade}(\partial_{b}\partial_{c}X^{i}_{\alpha})(\partial_{e}\partial_{d}X^{i}_{\beta})=\nonumber \\
&&\Big(\Psi^{\dag a,U}\langle
    F_{abc},F^{a'bc}\rangle\Psi_{a'}^{L}+\Psi^{\dag a,L}\langle
    F_{abc},F^{a'bc}\rangle\Psi_{a'}^{U}-\nonumber \\
    &&\Psi^{\dag a,U}\langle
    F_{abc},F^{a'bc}\rangle\Psi_{a'}^{U}-\Psi^{\dag a,L}\langle
    F_{abc},F^{a'bc}\rangle\Psi_{a'}^{L}+\nonumber \\
    &&\Psi^{\dag a,L}\Psi^{\dag d,U}\partial_{d}\partial^{d'}\langle
    F_{abc},F^{a'bc}\rangle\Psi_{a'}^{L}\Psi^{U}_{d'}-\nonumber \\
    &&\Psi^{\dag a,L}\Psi^{\dag d,U}\partial_{d}\langle
    F_{abc},F^{a'bc}\rangle\Psi_{a'}^{L}-\nonumber \\
    &&\Psi^{\dag a,U}\Psi^{\dag d,L}\partial_{d}\langle
    F_{abc},F^{a'bc}\rangle\Psi_{a'}^{U}\Big)+\nonumber \\
          &&  \Big(\psi^{\dag i,U}\langle
              F_{ijk},F^{i'jk}\rangle\psi_{i'}^{L}+\psi^{\dag i,L}\langle
              F_{ijk},F^{i'jk}\rangle\psi_{i'}^{U}-\nonumber \\
              &&\psi^{\dag i,U}\langle
              F_{ijk},F^{i'jk}\rangle\psi_{i'}^{U}-\psi^{\dag i,L}\langle
              F_{ijk},F^{i'jk}\rangle\psi_{i'}^{L}+\nonumber \\
              &&\psi^{\dag i,L}\psi^{\dag m,U}\partial_{m}\partial^{m'}\langle
              F_{ijk},F^{i'jk}\rangle\psi_{i'}^{L}\psi^{U}_{m'}-\nonumber \\
              &&\psi^{\dag i,L}\psi^{\dag m,U}\partial_{m}\langle
              F_{ijk},F^{i'jk}\rangle\psi_{i'}^{L}-\nonumber \\
              &&\psi^{\dag i,U}\psi^{\dag m,L}\partial_{m}\langle
              F_{ijk},F^{i'jk}\rangle\psi_{i'}^{U}\Big)+\nonumber \\
                    && \Big(\Psi^{\dag a,U}\langle
                        F_{abc},F^{i'bc}\rangle\psi_{i'}^{L}+\Psi^{\dag a,L}\langle
                        F_{abc},F^{i'bc}\rangle\psi_{i'}^{U}-\nonumber \\
                        &&\Psi^{\dag a,U}\langle
                        F_{abc},F^{i'bc}\rangle\psi_{i'}^{U}-\Psi^{\dag a,L}\langle
                        F_{abc},F^{i'bc}\rangle\psi_{i'}^{L}+\nonumber \\
                        &&\Psi^{\dag a,L}\Psi^{\dag d,U}\partial_{d}\partial^{i'}\langle
                        F_{abc},F^{j'bc}\rangle\psi_{j'}^{L}\psi^{U}_{i'}-\nonumber \\
                        &&\Psi^{\dag a,L}\Psi^{\dag d,U}\partial_{d}\langle
                        F_{abc},F^{i'bc}\rangle\psi_{i'}^{L}-\nonumber \\
                        &&\Psi^{\dag a,U}\Psi^{\dag d,L}\partial_{d}\langle
                        F_{abc},F^{i'bc}\rangle\psi_{i'}^{U}\Big) \nonumber \\
&&\nonumber \\
&&\nonumber \\
&&(\psi^{U}_{a})^{2}=0,\;(\psi^{L}_{a})^{2}=0\Rightarrow F(X)=\Sigma_{j}X_{j}^{2}=\Sigma_{j}[\psi^{U}_{a}A^{ab,j}\psi^{L}_{b}-\psi^{L}_{a}A_{ab,j}\psi^{U}_{b}]^{2}=\nonumber \\
&&\Sigma_{j}(\psi^{U}_{a})^{2}(A^{ab,j})^{2}(\psi^{L}_{b})^{2}+(\psi^{L}_{a})^{2}(A_{ab,j})^{2}(\psi^{U}_{b})^{2}-(\psi^{U}_{a}A^{ab,j}\psi^{L}_{b})(\psi^{L}_{a}A_{ab,j}\psi^{U}_{b})\nonumber \\
&&-(\psi^{L}_{a}A_{ab,j}\psi^{U}_{b})(\psi^{U}_{a}A^{ab,j}\psi^{L}_{b})=0\nonumber \\
&&\nonumber \\
&&\nonumber \\
&&(\psi^{U}_{a})^{2}=0,\;(\psi^{L}_{a})^{2}=0\Rightarrow \langle[X^{k},X^{i},X^{j}],[X_{k},X_{i},X_{j}]\rangle=\Sigma_{n}\Sigma_{m}\alpha_{n+m}(\psi^{L})^{2n}(\psi^{U})^{2m}=0\label{s7}
\end{eqnarray}

 where a,b,c are indices of bound electrons and i,j,k are indices of free electrons. In above equation, all terms in action can be obtained in terms of couplings between parallel spins and anti-parallel spins.  It is clear from above calculations that by joining electrons and formation a pair,  two form gauge fields are created that
 play the role of tensor mode of graviton in a graphene ($\Psi^{\dag a,U}\langle
                         F_{abc},F^{i'bc}\rangle\psi_{i'}^{L}$). In fact, the reason for the emergence of attraction force between electrons with anti-parallel spins is the production of gravitational force between them.

                           Previously in \cite{q6,q7,qq8}, it has been  proposed a new model in M-theory which allows for constructing all Mp-branes from M1 and M0-branes. In this model, it has been shown that  for M0-branes, there is no guage field and only scalars have attached to it, however by joining M0-branes and formation of M1-branes, gauge fields emerges on it. These M1-branes are linked to anti-M1-branes and form a new system. For this system, metric can be constructed from metrics of two M1's as follows:

                          $\textbf{Metric of system}=(\textbf{Metric M1})_{1}\otimes (\textbf{Metric M1})_{2}-(\textbf{Metric M1})_{2}\otimes (\textbf{Metric M1})_{1}$

                          Thus metric of system can be antisymmetric. On the other hand, tensor mode of graviton has a direct relation with metric  and can be anti-symmetric.  We have the same conditions for graphene. In a graphene, the metric can be constructed from the metric of pairs such as for two electrons of a pair: $\textbf{Metric of pair}=(\textbf{Metric electron})_{1}\otimes (\textbf{Metric electron})_{2}-(\textbf{Metric electron})_{2}\otimes (\textbf{Metric electron})_{1}$. This metric can be antisymmetric and thus tensor mode of graviton may be antisymmetric. For this reason, two form gauge fields have a direct relation with graviton and also with metric of pairs and the graphene. To clear the meaning of metric of pair and electron, it's needed that we explain more. In a graphene, each electron interacts with other electrons and experiences different forces from their sides. These forces leads to the emergence of the tensor of energy-momentum and consequently producing an special curvature around this electron. Consequently, this curvature creates an special metric for each electron. On the other hand, interactions of one electron may be different from other one as due to difference in spin and other properties and thus, each electron has it's special metric. When two electrons join to each other and form a system, both of them undergo the same forces from the side of other electrons and consequently experience the same curvature and metric which is known as the metric of pair. Each pair has different properties from other pairs. For example, some pairs are constructed from joining two free electrons, some others are created from linking one bound electron to atom and one free electron and some are produced by two bound electrons. Each of these pairs has it's special curvature and metric and thus for obtaining total curvature and metric  of system, we should sum over curvatures and metrics of all of these pairs. As a result, the metric of system is obtained by the average over the metrics of all pairs in system and the metric of each pair can be constructed from the metric of each electron in it.   Using this method in a graphene, we
 can obtain the relation between fermions and curvatures \cite {q6,q7,q8,qq8,q9,q10,q11,q12,m23,m24,m25}:

 \begin{eqnarray}
 &&A^{ab}=g^{ab}=h^{ab}=h_{1}^{ab}\otimes h_{2}^{ab}-h_{2}^{ab}\otimes h_{1}^{ab} ~ ~  and  ~ ~ a,b,c=\mu,\nu,\lambda
 \Rightarrow \nonumber\\&& F_{abc}=\partial_{a} A_{bc}-\partial_{b}
 A_{ca}+\partial_{c}
 A_{ab}=2(\partial_{\mu}g_{\nu\lambda}+\partial_{\nu}g_{\mu\lambda}-\partial_{\lambda}g_{\mu\nu})=
 2\Gamma_{\mu\nu\lambda}\nonumber\\&&\nonumber\\&&\langle
 F^{\rho}\smallskip_{\sigma\lambda},F^{\lambda}\smallskip_{\mu\nu}\rangle=
 \langle[X^{\rho},X_{\sigma},X_{\lambda}],[X^{\lambda},X_{\mu},X_{\nu}]\rangle=\nonumber\\&&
 [X_{\nu},[X^{\rho},X_{\sigma},X_{\mu}]]-[X_{\mu},[X^{\rho},X_{\sigma},X_{\nu}]]+[X^{\rho},X_{\lambda},X_{\nu}]
 [X^{\lambda},X_{\sigma},X_{\mu}]
 -[X^{\rho},X_{\lambda},X_{\mu}][X^{\lambda},X_{\sigma},X_{\nu}]=\nonumber\\&&\partial_{\nu}
 \Gamma^{\rho}_{\sigma\mu}-\partial_{\mu}\Gamma^{\rho}_{\sigma\nu}+\Gamma^{\rho}_{\lambda\nu}
 \Gamma^{\lambda}_{\sigma\mu}-\Gamma^{\rho}_{\lambda\mu}\Gamma^{\lambda}_{\sigma\nu}
 =R^{\rho}_{\sigma\mu\nu}\nonumber\\&&\langle
     F_{abc},F_{a'}^{bc}\rangle=R_{aa'}\nonumber\\&& R_{MN}=R_{aa'}+R_{ia'}+R_{ij'}=R_{Free-Free}+R_{Free-Bound}+R_{Bound-Bound}\label{s8}
 \end{eqnarray}

 This equation shows that two form gauge fields play the role of graviton in a graphene system. These gravitons are exchanged between electrons and produces three types of curvatures. One curvature is created as due to exchanging graviton between two free electrons, other is produced as due to exchanging graviton between free and bound electrons and third types is emerged by exchanging graviton between two bound electrons.

 Until now, we have found that $\langle
      F_{abc},F_{a'}^{bc}\rangle $ in action (\ref{s5} and \ref{ss5}), can be replaced by different types of curvatures ($R_{aa'}$). We like to obtain the relation between other terms in these actions like ($\langle \partial^{b}\partial^{a}X^{i},\partial_{b}\partial_{a}X^{i}\rangle$) and curvatures. This helps us to write action in terms of only curvatures and by solving the equation of motion, obtain the best place for defects in a graphene which leads to an increase in curvature and action of system. This curvature has a direction relation with tensor of energy-momentum and also momentum has a direct relation with force. By increasing the curvature, the applied force to electrons becomes stronger and electron moves in one special direction which this leads to an increase in conductivity of system. To achieve this aim, we use of  ($X\rightarrow \psi^{U}_{a}A^{ab}\psi^{L}_{b}-\psi^{L}_{a}A_{ab}\psi^{U}_{b}$) and equations (\ref{s7} and \ref{s8} ) in  $\kappa^{\mu}_{\nu}$ which includes remaining terms ($\langle \partial^{b}\partial^{a}X^{i},\partial_{b}\partial_{a}X^{i}\rangle$) in action (\ref{s5} and \ref{ss5}). In fact,
      in this method, first, we  replace ($\langle \partial^{b}\partial^{a}X^{i},\partial_{b}\partial_{a}X^{i}\rangle$) by results of equation (\ref{s7}) in $\kappa^{\mu}_{\nu}$ and write it in terms of gauge fields and spinors. Second, using equation (\ref{s6}), we replace spinors by gauge fields and write ($\langle \partial^{b}\partial^{a}X^{i},\partial_{b}\partial_{a}X^{i}\rangle$) in terms of $\langle
            F_{abc},F_{a'}^{bc}\rangle $. Finally, using results of equation (\ref{s8}), we replace ( $\langle
            F_{abc},F_{a'}^{bc}\rangle $) by curvatures.   We obtain:

 \begin{eqnarray}
 &&\kappa^{\mu}_{\nu}=\delta^{\mu}_{\nu}-\sqrt{\delta^{\mu}_{\nu}-H^{\mu}_{\nu}}\nonumber\\
 && H= g^{\mu\nu}H_{\mu\nu}=
 \delta^{\mu}_{\nu}-\langle \partial^{b}\partial^{a}X^{i},\partial_{b}\partial_{a}X^{i}\rangle =\nonumber\\
 && \delta^{\mu}_{\nu}+ \Big(\Psi^{\dag a,U}R_{aa'} \Psi^{a',L}+\Psi^{\dag a,L}R_{aa'}\Psi^{a',U}-\nonumber \\
      &&\Psi^{\dag a,U}R_{aa'}\Psi^{a',U}-\Psi^{\dag a,L}R_{aa'} \Psi^{a',L}+\nonumber \\
      &&\Psi^{\dag a,L}\Psi^{\dag d,U}\partial_{d}\partial^{d'}R_{aa'}\Psi^{a',L}\Psi^{U}_{d'}-\nonumber \\
      &&\Psi^{\dag a,L}\Psi^{\dag d,U}\partial_{d}R_{aa'} \Psi^{a',L}-\nonumber \\
      &&\Psi^{\dag a,U}\Psi^{\dag d,L}\partial_{d}R_{aa'}\Psi^{a',U}\Big)+\nonumber \\
            &&  \Big(\psi^{\dag i,U}R_{ii'} \psi^{i',L}+\psi^{\dag i,L}R_{ii'}\psi^{i',U}-\nonumber \\
                &&\Big(\psi^{\dag i,U}R_{ii'}\psi^{i',U}-\psi^{\dag i,L}R_{ii'}\psi^{i',L}+\nonumber \\
                &&\psi^{\dag i,L}\psi^{\dag m,U}\partial_{m}\partial^{m'}R_{ii'}\psi^{i',L}\psi^{U}_{m'}-\nonumber \\
                &&\psi^{\dag i,L}\psi^{\dag m,U}\partial_{m}R_{ii'}\psi^{i',L}-\nonumber \\
                &&\psi^{\dag i,U}\psi^{\dag m,L}\partial_{m}R_{ii'}\psi^{i',U}\Big)+\nonumber \\
                      && \Big(\Psi^{\dag a,U}R_{ai'}\psi^{i',L}+\Psi^{\dag a,L}R_{ai'}\rangle\psi^{i',U}-\nonumber \\
                          &&\Psi^{\dag a,U}R_{ai'}\psi^{i',U}-\Psi^{\dag a,L}R_{ai'}\psi^{i',L}+\nonumber \\
                          &&\Psi^{\dag a,L}\Psi^{\dag d,U}\partial_{d}\partial^{i'}R_{ai'}\psi_{j'}^{L}\psi^{i',U}-\nonumber \\
                          &&\Psi^{\dag a,L}\Psi^{\dag d,U}\partial_{d}R_{ai'}\psi^{i',L}-\nonumber \\
                          &&\Psi^{\dag a,U}\Psi^{\dag d,L}\partial_{d}R_{ai'}\psi^{i',U}\Big)\approx \nonumber \\
                                                    && R_{Free electrons}^{2}+R_{Free-Bound}^{2}+R_{Bound}^{2}+\nonumber \\&& ( R_{Free electrons}^{2}+R_{Free-Bound}^{2}+R_{Bound}^{2})\partial^{2}( R_{Free electrons}+R_{Free-Bound}+R_{Bound})\label{s9}
 \end{eqnarray}
  We show that $\kappa^{\mu}_{\nu}$ can be written in terms of curvatures only. On the hand, curvature depends on the $\Gamma$ ($R^{\rho}_{\sigma\mu\nu}=\partial_{\nu}
              \Gamma^{\rho}_{\sigma\mu}-\partial_{\mu}\Gamma^{\rho}_{\sigma\nu}+\Gamma^{\rho}_{\lambda\nu}
              \Gamma^{\lambda}_{\sigma\mu}-\Gamma^{\rho}_{\lambda\mu}\Gamma^{\lambda}_{\sigma\nu}$) (see(\ref{s8})) and $\Gamma$ depends on the metric ($\Gamma_{\mu\nu\lambda}=(\partial_{\mu}g_{\nu\lambda}+\partial_{\nu}g_{\mu\lambda}-\partial_{\lambda}g_{\mu\nu})$) and  thus, $\kappa^{\mu}_{\nu}$  depends on the metric so.  These calculations show that there are three types of curvatures. one type of them are produced by coupling of free electrons to each other. Another type is emerges as due to coupling of free electrons to bound electrons and final type which is created by coupling of bound electrons. If curvature be between
 anti-symmetric fermion, it will be positive and if coupling be between parallel spins, the curvature will be negative. In fact, this equation shows that the gravity between anti-parallel spins ($\Psi^{\dag a,U}R_{aa'} \Psi^{a',L}$) produces the attracting force and anti-gravity between parallel spins (-$\Psi^{\dag a,U}R_{aa'} \Psi^{a',U}$) creates the repelling force.  In previous studies, it has been
 shown that  $\kappa$ has a direct relation with curvature scalars
 ($R$) \cite{m23,m24,m25}. Our calculations give the same results. As you can see in equation (\ref{s9}), $\kappa$ is written in terms of ($\langle \partial^{b}\partial^{a}X^{i},\partial_{b}\partial_{a}X^{i}\rangle$).  On the other hand, in equation (\ref{s7}), it has been asserted that  ($\langle \partial^{b}\partial^{a}X^{i},\partial_{b}\partial_{a}X^{i}\rangle$) can be given in terms of  ( $\langle
              F_{abc},F_{a'}^{bc}\rangle $) and spinors. The relation between spinors and ( $\langle
                            F_{abc},F_{a'}^{bc}\rangle $) has been obtained in (\ref{s6}) and thus, we can replace spinors by gauge fields.  Also, in (\ref{s8}), it has been shown that ( $\langle
                            F_{abc},F_{a'}^{bc}\rangle $) can be written in terms of curvatures. Thus, in equation (\ref{s9}), we obtain the relation between ($\langle \partial^{b}\partial^{a}X^{i},\partial_{b}\partial_{a}X^{i}\rangle$) and $\kappa^{\mu}_{\rho}$ in terms of curvatures. However $\kappa^{\mu}_{\rho}$ depends on the square of ($\sqrt{\langle \partial^{b}\partial^{a}X^{i},\partial_{b}\partial_{a}X^{i}\rangle}$), while in action we need to ($\langle \partial^{b}\partial^{a}X^{i},\partial_{b}\partial_{a}X^{i}\rangle$) and for this reason we use of following relation :

 \begin{eqnarray}
 &&\delta^{\rho\sigma}_{\mu\nu}\kappa^{\mu}_{\rho}\kappa^{\nu}_{\sigma}= R_{Free-Free}^{2}+R_{Free-Bound}^{2}+R_{Bound-Bound}^{2}+\nonumber \\&& ( R_{Free-Free}^{2}+R_{Free-Bound}^{2}+R_{Bound-Bound}^{2})\partial^{2}( R_{Free-Free}+R_{Free-Bound}+R_{Bound-Bound})
 \label{s10}
 \end{eqnarray}

This result that scalar curvatures like tensor curvatures can be written in terms of couplings of free and bound electrons. For couplings of anti-parallel spins, curvature is positive and for couplings of parallel spins, curvature is negative. Now, we can obtain the form of the relevant action of conductivity  in equation (\ref{ss5}) in
 terms of curvature scalars and tensors in gravity. The method is similar to \cite{q7,qq8}. We begin with the Lagrangian of one point in a graphene which is a part of action  in equation (\ref{ss5}):

 \begin{eqnarray}
 &&
 L=\delta^{a_{1},a_{2}...a_{n}}_{b_{1}b_{2}....b_{n}}L^{b_{1}}_{a_{1}}...L^{b_{n}}_{a_{n}}
 \quad a,b,c=\mu,\nu,\lambda \nonumber\\\nonumber\\
 && L^{b}_{a}=\delta_{a}^{b}\det (P_{abc}[ E_{mnl} +E_{mij}(Q^{-1}-\delta)^{ijk}E_{kln}]+
 \lambda F_{abc})\nonumber\\&&\nonumber\\
 && L^{b}_{a}=\delta_{a}^{b}\det\left(P_{abc}[ E_{mnl}
 +E_{mij}(Q^{-1}-\delta)^{ijk}E_{kln}]\right)+\delta_{a}^{b}\lambda^{2}\det(F)\label{s11}\end{eqnarray}

 Using the above equation, we can obtain the relevant terms of determinant
 in action (\ref{ss5}) separately. Substituting equations (\ref{s6},\ref{s7},\ref{s8},\ref{s9},\ref{s10})
 in determinants (\ref{s11}), we obtain:

 \begin{eqnarray}
 \det(F)=\delta_{\rho\sigma}^{\mu\nu}\langle
 F^{\rho\sigma}\smallskip_{\lambda},F^{\lambda}\smallskip_{\mu\nu}\rangle
 =\delta_{\rho\sigma}^{\mu\nu}(R^{\rho\sigma}_{Free-Free,\mu\nu}+R^{\rho\sigma}_{Bound-Bound,\mu\nu}+R^{\rho\sigma}_{Free-Bound,\mu\nu})\label{s12}\end{eqnarray}

 \begin{eqnarray}
 && \det(P_{abc}[ E_{mnl}
 +E_{mij}(Q^{-1}-\delta)^{ijk}E_{kln}])=\nonumber\\&&\delta_{\rho\sigma}^{\mu\nu}
 \Biggl[(g^{\mu}_{\rho}g^{\nu}_{\sigma}+ \langle \partial^{b}\partial^{a}X^{i},\partial_{b}\partial_{a}X^{i}\rangle+..)-
 \nonumber\\&&\frac{(g^{\mu}_{\rho}g^{\nu}_{\sigma}+
 \langle \partial^{b}\partial^{a}X^{i},\partial_{b}\partial_{a}X^{i}\rangle+...)}{[(\lambda)^{2}
 \det([X^{j}_{\alpha}T^{\alpha},X^{k}_{\beta}T^{\beta},X^{k'}_{\gamma}T^{\gamma}])]}\Biggr]=
 \nonumber\\&& [R_{Free-Free}^{2}+R_{Free-Bound}^{2}+R_{Bound-Bound}^{2}+\nonumber \\&& ( R_{Free electrons}^{2}+R_{Free-Bound}^{2}+R_{Bound}^{2})\partial^{2}( R_{Free-Free}+R_{Free-Bound}+R_{Bound-Bound})]\times\nonumber \\&&
 \left(1-\frac{1}{\left[(\lambda)^{2}\det([X^{j}_{\alpha}T^{\alpha},X^{k}_{\beta}T^{\beta},X^{k'}_{\gamma}T^{\gamma}])\right]}\right)=
 \nonumber\\&&[R_{Free-Free}^{2}+R_{Free-Bound}^{2}+R_{Bound-Bound}^{2}+\nonumber \\&& ( R_{Free-Free}^{2}+R_{Free-Bound}^{2}+R_{Bound-Bound}^{2})\partial^{2}( R_{Free electrons}+R_{Free-Bound}+R_{Bound})](1-\frac{1}{m_{g}^{2}})\label{s13}
 \end{eqnarray}

 where
 $m_{g}^{2}=[(\lambda)^{2}\det([X^{j}_{\alpha}T^{\alpha},X^{k}_{\beta}T^{\beta},X^{k'}_{\gamma}T^{\gamma}])]
 $ is the square of gauge mass. Dimension of $X^{j}$ is of order of length and dimension of $\lambda^{2}$ is of order of $\frac{1}{l_{a}^{3}}$ where $l_{a}$ is the relative distance between two atoms and thus, $m_{g}^{2}$ is dimensionless. Using this
 definition, we can obtain another term of the determinant:

 \begin{eqnarray}
  &&\det(Q)\sim
 (i)^{2}(\lambda)^{2}\det([X^{j}_{\alpha}T^{\alpha},X^{k}_{\beta}T^{\beta},X^{k'}_{\gamma}T^{\gamma}])\det(E)\sim
 \nonumber\\&&
 -[(\lambda)^{2}\det([X^{j}_{\alpha}T^{\alpha},X^{k}_{\beta}T^{\beta},X^{k'}_{\gamma}T^{\gamma}])]\det(g)=-m_{g}^{2}\det(g)
 \label{s14}
 \end{eqnarray}

 By substituting equations (\ref{s9}) ,(\ref{s10}), (\ref{s11}),
 (\ref{s12}) ,(\ref{s13}), and (\ref{s14}) into the action
 (\ref{ss5}), we get:

 \begin{eqnarray}
 && L_{b}^{a}=\delta_{b}^{a}\Bigl[\Big(-(1-m_{g}^{2})[R_{Free-Free}^{2}+R_{Free-Bound}^{2}+R_{Bound-Bound}^{2}+\nonumber \\&& ( R_{Free-Free}^{2}+R_{Free-Bound}^{2}+R_{Bound-Bound}^{2})\partial^{2}( R_{Free-Free}+R_{Free-Bound}+R_{Bound-Bound})]\nonumber\\&&+
 m_{g}^{2}\lambda^{2}\delta_{\rho\sigma}^{\mu\nu}(R^{\rho\sigma}_{Free-Free,\mu\nu}+R^{\rho\sigma}_{Bound-Bound,\mu\nu}+R^{\rho\sigma}_{Free-Bound,\mu\nu})\Big)\Bigr]
 \label{s15}
 \end{eqnarray}

  This Lagrangian shows that the for calculating the conductivity in a graphene, we should calculate the sum over curvatures produced by free and bound electrons at each point of graphene. This is because that curvature has a direct relation with energy momentum tensor and if curvature at one point be zero, electrons can't continue to it's motion and conductivity at that point disappears. In fact, curvatures determine the conductivity of graphene and motion of electrons in a certain direction. Now, we can obtain the relevant action for conductivity in graphene by using the
 method in \cite{q6,q7,qq8,q8} and substituting equation (\ref{s15}) in equation (\ref{ss5}) :

 \begin{eqnarray}
 && S_{co-Graphene} = - \int d^{4}x \L_{Mp}  \nonumber\\&& \L_{Mp}=\det(M)
 \quad L_{M1,i}=L^{b_{i}}_{a_{i}}=\det(M_{i})\sim M_{i}~~~
 \mbox{where,}
 ~~~~\det(M)=\sum_{n=1}^{p}\delta^{a_{1},a_{2}...a_{n}}_{b_{1}b_{2}....b_{n}}M^{b_{1}}_{a_{1}}...M^{b_{n}}_{a_{n}}\Rightarrow
 \nonumber\\&&
 \L_{Mp}=\det(M)=\sum_{n=1}^{p}\delta^{a_{1},a_{2}...a_{n}}_{b_{1}b_{2}....b_{n}}L^{b_{1}}_{a_{1}}...L^{b_{n}}_{a_{n}}
 \nonumber~~~ \mbox{where,}
 ~~~~\delta^{a_{1},a_{2}...a_{n}}_{b_{1}b_{2}....b_{n}}
 \delta^{\rho_{1}\sigma_{1}}_{\mu_{1}\nu_{1}}...\delta^{\rho_{p}\sigma_{p}}_{\mu_{p}\nu_{p}}
 =\delta^{\rho_{1}\sigma_{1}...\rho_{p}\sigma_{p}}_{\mu_{1}\nu_{1}...\mu_{p}\nu_{p}}\nonumber\\&&
 \sqrt{-g}=\sqrt{-\det(g)}=\sqrt{-\det(g_{1}g_{2}...g_{p})}=\sqrt{-\det(g_{1})\det(g_{2})...\det(g_{p})}\label{s16}
 \end{eqnarray}

 so we obtain

 \begin{eqnarray}
 S_{co-Graphene} &=& -\int d^{4}x \sum_{n=1}^{Q}
 \delta^{a_{1},a_{2}...a_{n}}_{b_{1}b_{2}....b_{n}}L^{b_{1}}_{a_{1}}...L^{b_{p}}_{a_{p}}=\nonumber\\&&
 - \int d^{4}x \sum_{n=1}^{Q}
 \delta^{a_{1},a_{2}...a_{n}}_{b_{1}b_{2}....b_{n}}\nonumber\\
 &&\Big( \Bigl[\Big(-(1-m_{g}^{2})[R_{Free-Free}^{2}+R_{Free-Bound}^{2}+R_{Bound-Bound}^{2}+\nonumber \\&& ( R_{Free-Free}^{2}+R_{Free-Bound}^{2}+R_{Bound-Bound}^{2})\partial^{2}( R_{Free-Free}+R_{Free-Bound}+R_{Bound-Bound})]\nonumber\\&&+
  m_{g}^{2}\lambda^{2}\delta_{\rho_{1}\sigma_{1}}^{\mu_{1}\nu_{1}}(R^{\rho_{1}\sigma_{1}}_{Free-Free,\mu_{1}\nu_{1}}+R^{\rho_{1}\sigma_{1}}_{Bound-Bound,\mu_{1}\nu_{1}}+R^{\rho_{1}\sigma_{1}}_{Free-Bound,\mu_{1}\nu_{1}})\Big)\Bigr]
 ^{b_{1}}_{a_{1}}\times
 \nonumber\\&&.......\times\nonumber\\
 && \Bigl[\Big(-(1-m_{g}^{2})[R_{Free-Free}^{2}+R_{Free-Bound}^{2}+R_{Bound-Bound}^{2}+\nonumber \\&& ( R_{Free-Free}^{2}+R_{Free-Bound}^{2}+R_{Bound-Bound}^{2})\partial^{2}( R_{Free-Free}+R_{Free-Bound}+R_{Bound-Bound})]\nonumber\\&&+
  m_{g}^{2}\lambda^{2}\delta_{\rho_{n}\sigma_{n}}^{\mu_{n}\nu_{n}}(R^{\rho_{n}\sigma_{n}}_{Free-Free,\mu_{n}\nu_{n}}+R^{\rho_{n}\sigma_{n}}_{Bound-Bound,\mu_{n}\nu_{n}}+R^{\rho_{n}\sigma_{n}}_{Free-Bound,\mu_{n}\nu_{n}})\Big)\Bigr]
 ^{b_{n}}_{a_{n}}\Big)^{\frac{1}{2}}\nonumber\\&=&- \int d^{4}x
 \Biggl[\sqrt{-g}\Bigl(\sum_{n=1}^{Q}\Big(-(1-m_{g}^{2})^{n}
 [R_{Free-Free}^{2}+R_{Free-Bound}^{2}+R_{Bound-Bound}^{2}+\nonumber \\&& ( R_{Free-Free}^{2}+R_{Free-Bound}^{2}+R_{Bound-Bound}^{2})\partial^{2}( R_{Free-Free}+R_{Free-Bound}+R_{Bound-Bound})]^{n}+\nonumber\\&&
 \sum_{n=1}^{Q}m_{g}^{2n}\lambda^{2n}\delta^{\rho_{1}\sigma_{1}...\rho_{n}\sigma_{n}}_{\mu_{1}\nu_{1}...\mu_{n}\nu_{n}}
 (R^{\rho_{1}\sigma_{1}}_{Free-Free,\mu_{1}\nu_{1}}+R^{\rho_{1}\sigma_{1}}_{Bound-Bound,\mu_{1}\nu_{1}}+R^{\rho_{1}\sigma_{1}}_{Free-Bound,\mu_{1}\nu_{1}})...\times\nonumber\\&&(R^{\rho_{n}\sigma_{n}}_{Free-Free,\mu_{n}\nu_{n}}+R^{\rho_{n}\sigma_{n}}_{Bound-Bound,\mu_{n}\nu_{n}}+R^{\rho_{n}\sigma_{n}}_{Free-Bound,\mu_{n}\nu_{n}})
 \Bigr)\Biggr]\Big)^{\frac{1}{2}} \label{s17}
 \end{eqnarray}

  This equation shows that the conductivity of a graphene depends on the curvatures that are produced by the couplings of free electrons, bound electrons and also interaction of free electrons by bound electrons at each atom. This is because that curvature has a direct relation with tensor of energy-momentum. When curvature of a system is not zero, the applied momentum on the electrons is not zero and electrons move in an special direction. This leads to creation of conductivity in a graphene.  We have:

  \begin{eqnarray}
  &&  \delta^{\rho_{1}\sigma_{1}...\rho_{n}\sigma_{n}}_{\mu_{1}\nu_{1}...\mu_{n}\nu_{n}}
  R_{\rho_{1}\sigma_{1}}^{\mu_{1}\nu_{1}}...R_{\rho_{n}\sigma_{n}}^{\mu_{n}\nu_{n}}\rightarrow
  R^{n}\nonumber\\&&\Rightarrow S_{co-Graphene}= -\int d^{4}x
  \Biggl[\sqrt{-g}\Big(\sum_{n=1}^{Q}\Big(\Big(-(1-m_{g}^{2})^{n}
   [R_{Free-Free}^{2}+R_{Free-Bound}^{2}+R_{Bound-Bound}^{2}+\nonumber \\&& ( R_{Free-Free}^{2}+R_{Free-Bound}^{2}+R_{Bound-Bound}^{2})\partial^{2}( R_{Free-Free}+R_{Free-Bound}+R_{Bound-Bound})]^{n}\nonumber\\&&+\lambda^{2n}m_{g}^{2n}\left[R_{Free-Free}+R_{Free-Bound}+R_{Bound-Bound}\right]^{n}\Big)\Biggr]\Big)^{\frac{1}{2}}\nonumber\\&&\Rightarrow
  S_{co-Graphene}= -\int d^{4}x
  \Biggl[\sqrt{-g}\Big(F(R)\Big)\Biggr]\nonumber\\&&
  F(R)=\sum_{n=1}^{Q}\Big(-(1-m_{g}^{2})^{n}
     [R_{Free-Free}^{2}+R_{Free-Bound}^{2}+R_{Bound-Bound}^{2}+\nonumber \\&& ( R_{Free-Free}^{2}+R_{Free-Bound}^{2}+R_{Bound-Bound}^{2})\partial^{2}( R_{Free-Free}+R_{Free-Bound}+R_{Bound-Bound})]^{n}\nonumber\\&&+\lambda^{2n}m_{g}^{2n}\left[R_{Free-Free}+R_{Free-Bound}+R_{Bound-Bound}\right]^{n}\Big)^{\frac{1}{2}}\label{s18}
  \end{eqnarray}

 In this equation, $F(R)$-gravity has been written in terms of different curvatures which are produced by interactions of 1.two bound electrons, 2. bound and free electrons, 3. two free electrons of an atom. Previously in equation (\ref{s8}), we have shown that curvature has a direct relation with ($\langle
  F^{\rho}\smallskip_{\sigma\lambda},F^{\lambda}\smallskip_{\mu\nu}\rangle$). Also, in this equation, we show that $F^{\rho}\smallskip_{\sigma\lambda}=\partial^{\rho}A_{\sigma\lambda}+..$ depends on two form gauge field $A_{\sigma\lambda}$. On the other hand, in this equation and also in explainations before it, we discuss about the relation of this gauge field and the  metric of pairs ($A_{\sigma\lambda}\approx g_{\sigma\lambda}$). Thus, the above action depends on the metric of pairs which are produced by interaction of two free electrons, two bound electrons and one free and one bound electron which are moving around one atom and then by summing over all Q atoms of graphene, we regard metrics of all pairs. In a graphene, we don't speak of the metric of one electron itself, because the curvatures produced by interact of electron with other electrons and we should speak about metric of pairs that are produced by these interactions and create curvatures. To clear this point, we should remind that curvature is related to ($\langle
    F^{\rho}\smallskip_{\sigma\lambda},F^{\lambda}\smallskip_{\mu\nu}\rangle$) and as you can see in equation (\ref{s6}), this expression depends on the wave functions of electrons and thus, curvatures depend on the wave functions of electrons in graphene system and in fact, interactions of all electrons have been regarded in calculations.  This equation shows that the exact form of $F(R)$ gravity in a graphene depends on the number of atoms and couplings of free and bound electrons at the defects. If identical spins couple to each other curvature is negative and if anti-parallel spins connect to each other, curvature is positive. For the symmetric graphene,  the number of coupling between identical fermions is equal to couplings on anti-symmetric fermions, total curvature becomes zero and conductivity disappears.  For graphene with defects, depend on the shape of the defect, the coupling of anti-symmetric spins increases and conductivity emerges.

  At this stage, we obtain the curvature produced by coupling of parallel spins and anti-parallel spins in terms of angle between sides of a defect. Graphene has a hexagonal shape and the  angle between neighbor atoms is 60. On the other hand, the spin of electrons that are located on neighbor atoms are in opposite directions respect to each other. Thus, for a symmetric graphene, the angle between anti-parallel spin is 60 (see Fig.1.) and the angle between parallel spins is 120 (see Fig.2.). In this system, the number of couplings between anti-symmetric spins and parallel spins are equal and curvature produced by anti-symmetric fermions are neutralized by symmetric fermions and total curvature is zero. However, for a defect (see Fig.3.), the separation distance between atoms changes and some new couplings are produced that generate a new curvature and conductivity. To obtain the wave equations from (\ref{s18}), we need to rewrite it in terms of coupling of parallel spins and also couplings of anti-parallel spins. Before, in equation (\ref{s8}), we have shown that curvature is related to ($\langle
    F^{\rho}\smallskip_{\sigma\lambda},F^{\lambda}\smallskip_{\mu\nu}\rangle$). Also, in equation (\ref{s6}), we have obtain this expression ($\langle
        F^{\rho}\smallskip_{\sigma\lambda},F^{\lambda}\smallskip_{\mu\nu}\rangle$) in terms of different derivatives of couplings of parallel spins ($\psi^{\dag a, U}\psi^{U}_{a},\psi^
         {\dag a, L}\psi^{L}_{a}$) and couplings of anti-parallel spins ($\psi^{\dag a, U}\psi^{L}_{a},\psi^{\dag a, L}\psi^{U}_{a}$). Thus, curvature can be written in terms of these couplings.  For simplicity,  we choose $l_{1}^{2}=\psi^{\dag a, U}\psi^{L}_{a}=\psi^{\dag a, L}\psi^{U}_{a}$, where $l_{1}^{2}$ is a symbol for couplings of anti-parallel spins and $l_{2}^{2}=\psi^{\dag a, U}\psi^{U}_{a}=\psi^
 {\dag a, L}\psi^{L}_{a}$ where $l_{2}^{2}$  is a symbol for couplings of parallel spins where these couplings ($l_{1}$ and $l_{2}$) only depends on the angles between sheets of  defects in graphene (For example, see the angle between sheets in Fig.3.). We obtain:

         \begin{figure*}[thbp]
         \includegraphics[width=6.0cm]{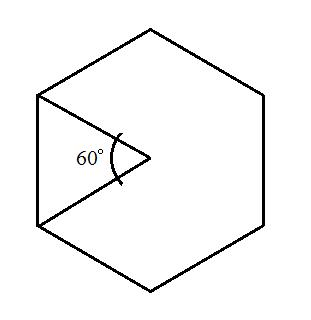}
         \caption{ The angle between neighbor atoms respect to center of hexagonal is 60.}
         \end{figure*}

           \begin{figure*}[thbp]
             \includegraphics[width=6.0cm]{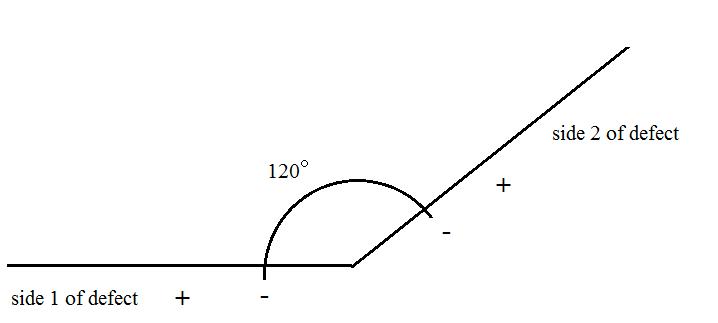}
             \caption{  In this graphene, the angle between sides of a defect is 120.}
           \end{figure*}

         \begin{figure*}[thbp]
         \includegraphics[width=6.0cm]{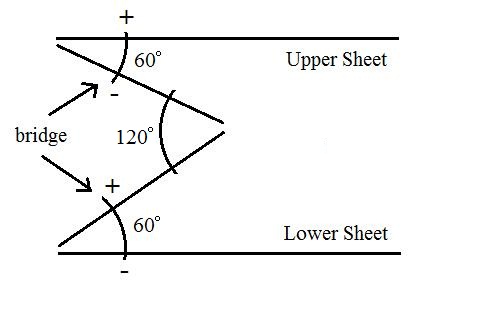}
         \caption{ In this graphene, the angle between graphene sheets and bridge and also between two sheets of bridge is 60.}
         \end{figure*}

    \begin{eqnarray}
    &&  S_{co-Geraphene}=V\int  d(\cos(\theta))\Sigma_{n=1}^{Q} \Big(m_{g}^{2}\lambda^{2}(2l_{1}^{2}-2l_{2}^{2}+2(l'_{1})^{2}-2(l'_{2})^{2}-(l'_{1})^{2}l_{2}-(l'_{2})^{2}l_{1})\nonumber\\&&-(1-m_{g}^{2})(2l_{1}^{2}-2l_{2}^{2}+2(l'_{1})^{2}-2(l'_{2})^{2}-(l'_{1})^{2}l_{2}-(l'_{2})^{2}l_{1})^{2}\times\nonumber\\&&[1+(2l_{1}^{2}-2l_{2}^{2}+2(l'_{1})^{2}-2(l'_{2})^{2}-(l'_{1})^{2}l_{2}-(l'_{2})^{2}l_{1})''] \Big)^{\frac{n}{2}}\label{s19}
    \end{eqnarray}

 where $'$ is the derivative respect to $d(\cos(\theta))$. The equation of motion for these equations are:

      \begin{eqnarray}
      &&\Big(\Big(m_{g}^{2}\lambda^{2}(4(l'_{1})-2(l'_{1})^{2})-(1-m_{g}^{2})(4(l'_{1})-2(l'_{1})l_{2})\times\nonumber\\&&(2l_{1}^{2}-2l_{2}^{2}+2(l'_{1})^{2}-2(l'_{2})^{2}-(l'_{1})^{2}l_{2}-(l'_{2})^{2}l_{1})\times\nonumber\\&&(2l_{1}^{2}-2l_{2}^{2}+2(l'_{1})^{2}-2(l'_{2})^{2}-(l'_{1})^{2}l_{2}-(l'_{2})^{2}l_{1})''+\nonumber\\&&
      (2(l''_{1})-2(l'_{1})l''_{2}-2(l''_{2}))(2l_{1}^{2}-2l_{2}^{2}+2(l'_{1})^{2}-2(l'_{2})^{2}-(l'_{1})^{2}l_{2}-(l'_{2})^{2}l_{1}) \Big)  \times\nonumber\\&& \Big(m_{g}^{2}\lambda^{2}(2l_{1}^{2}-2l_{2}^{2}+2(l'_{1})^{2}-2(l'_{2})^{2}-(l'_{1})^{2}l_{2}-(l'_{2})^{2}l_{1})\nonumber\\&&-(1-m_{g}^{2})(2l_{1}^{2}-2l_{2}^{2}+2(l'_{1})^{2}-2(l'_{2})^{2}-(l'_{1})^{2}l_{2}-(l'_{2})^{2}l_{1})^{2}\times\nonumber\\&&[1+(2l_{1}^{2}-2l_{2}^{2}+2(l'_{1})^{2}-2(l'_{2})^{2}-(l'_{1})^{2}l_{2}-(l'_{2})^{2}l_{1})'']\Big)^{\frac{n}{2}-1}\Big)'=\nonumber\\&&\Big(m_{g}^{2}\lambda^{2}(4l_{1}-(l'_{2})^{2})-(1-m_{g}^{2})(4l_{1}-(l'_{2})^{2})(2l_{1}^{2}-2l_{2}^{2}+2(l'_{1})^{2}-2(l'_{2})^{2}-(l'_{1})^{2}l_{2}-(l'_{2})^{2}l_{1})'' \Big)\times\nonumber\\&&\Big(m_{g}^{2}\lambda^{2}(2l_{1}^{2}-2l_{2}^{2}+2(l'_{1})^{2}-2(l'_{2})^{2}-(l'_{1})^{2}l_{2}-(l'_{2})^{2}l_{1})\nonumber\\&&-(1-m_{g}^{2})(2l_{1}^{2}-2l_{2}^{2}+2(l'_{1})^{2}-2(l'_{2})^{2}-(l'_{1})^{2}l_{2}-(l'_{2})^{2}l_{1})^{2}\times\nonumber\\&&[1+(2l_{1}
 ^{2}-2l_{2}^{2}+2(l'_{1})^{2}-2(l'_{2})^{2}-(l'_{1})^{2}l_{2}-(l'_{2})^{2}l_{1})'' ]\Big)^{\frac{n}{2}-1}\label{s20}
      \end{eqnarray}

            \begin{eqnarray}
            &&\Big(\Big(m_{g}^{2}\lambda^{2}(4(-l'_{2})+2(l'_{2})^{2})-(1-m_{g}^{2})(-4(l'_{2})+2(l'_{2})l_{1})\times\nonumber\\&&(2l_{1}^{2}-2l_{2}^{2}+2(l'_{1})^{2}-2(l'_{2})^{2}-(l'_{1})^{2}l_{2}-(l'_{2})^{2}l_{1})\times\nonumber\\&&(2l_{1}^{2}-2l_{2}^{2}+2(l'_{1})^{2}-2(l'_{2})^{2}-(l'_{1})^{2}l_{2}-(l'_{2})^{2}l_{1})''+\nonumber\\&&
            (-2(l''_{2})+2(l'_{2})l''_{1}+2(l''_{1}))(2l_{1}^{2}-2l_{2}^{2}+2(l'_{1})^{2}-2(l'_{2})^{2}-(l'_{1})^{2}l_{2}-(l'_{2})^{2}l_{1}) \Big)  \times\nonumber\\&& \Big(m_{g}^{2}\lambda^{2}(2l_{1}^{2}-2l_{2}^{2}+2(l'_{1})^{2}-2(l'_{2})^{2}-(l'_{1})^{2}l_{2}-(l'_{2})^{2}l_{1})\nonumber\\&&-(1-m_{g}^{2})(2l_{1}^{2}-2l_{2}^{2}+2(l'_{1})^{2}-2(l'_{2})^{2}-(l'_{1})^{2}l_{2}-(l'_{2})^{2}l_{1})^{2}\times\nonumber\\&& [1+(2l_{1}^{2}-2l_{2}^{2}+2(l'_{1})^{2}-2(l'_{2})^{2}-(l'_{1})^{2}l_{2}-(l'_{2})^{2}l_{1})''] \Big)^{\frac{n}{2}-1}\Big)'=\nonumber\\&&\Big(m_{g}^{2}\lambda^{2}(-4l_{2}+(l'_{1})^{2})-(1-m_{g}^{2})(-4l_{2}+(l'_{1})^{2})(-2l_{2}^{2}+2l_{1}^{2}-2(l'_{2})^{2}+2(l'_{1})^{2}+(l'_{2})^{2}l_{1}+(l'_{1})^{2}l_{2})'' \Big)\times\nonumber\\&&\Big(m_{g}^{2}\lambda^{2}(2l_{1}^{2}-2l_{2}^{2}+2(l'_{1})^{2}-2(l'_{2})^{2}-(l'_{1})^{2}l_{2}-(l'_{2})^{2}l_{1})\nonumber\\&&-(1-m_{g}^{2})(2l_{1}^{2}-2l_{2}^{2}+2(l'_{1})^{2}-2(l'_{2})^{2}-(l'_{1})^{2}l_{2}-(l'_{2})^{2}l_{1})^{2}\times\nonumber\\
 &&[1+(2l_{1}^{2}-2l_{2}^{2}+2(l'_{1})^{2}-2(l'_{2})^{2}-(l'_{1})^{2}l_{2}-(l'_{2})^{2}l_{1})''] \Big)^{\frac{n}{2}-1}\label{s21}
            \end{eqnarray}

           Solving these nonlinear coupled equations are very hard, however there are some mathematical methodes for them. For example, we can follow these steps. 1.We  choose  $l_{1}=\Sigma_{n,m}(f(x)-f(x_{0,1}))^{n}(g(x)-g(x_{0,2}))^{m}$ and $l_{2}=\Sigma_{n',m'}(f(x)-f(x_{0,1}))^{n'}(g(x)-g(x_{0,2}))^{m'}$.  2. We put these solutions in nonlinear equation and separate terms with the same order of x in $\{\}$. 3. We also use of some approximations like Newton expression and also taylor expression for writing radicals and functions in terms of different orders of $x$ and $x_{0}$.  4. We put terms of the same order of x in seperated $\{\}$. 5. Each of expression in $\{\}$ should be zero and we obtain some simple equations for $x_{0}$, m,n,m' and n'. 6.By solving them, we can obtain the approximate solutions of these equations. Here we choose $x=\cos(\theta)$ and obtain $x_{0,1}=\frac{1}{2}-\frac{m_{g}^{2}}{2}$ and $x_{0,2}=-\frac{1}{2}-\frac{m_{g}^{2}}{2}$.  The solution of these equations are approximately:

                        \begin{eqnarray}
                        &&l_{1}\approx \Sigma_{n=1}^{Q}[\frac{2m_{g}^{2}\lambda^{2}+2^{2}m_{g}^{4}\lambda^{4}(1+m_{g}^{2}+2\cos(\theta))+..+2^{2n}m_{g}^{4n}\lambda^{4n}(1+m_{g}^{2}+2\cos(\theta))^{2n}}{2\cos(\theta)-1+m_{g}^{2}}]\nonumber\\&&l_{2}\approx \Sigma_{n=1}^{Q}[\frac{2m_{g}^{2}\lambda^{2}+2^{2}m_{g}^{4}\lambda^{4}(-1+m_{g}^{2}+2\cos(\theta))+...+2^{2n}m_{g}^{4n}\lambda^{4n}(-1+m_{g}^{2}+2\cos(\theta))^{2n}}{1+m_{g}^{2}+2\cos(\theta)}]\label{s22}
                        \end{eqnarray}

           \begin{figure*}[thbp]
             \includegraphics[width=5.5cm]{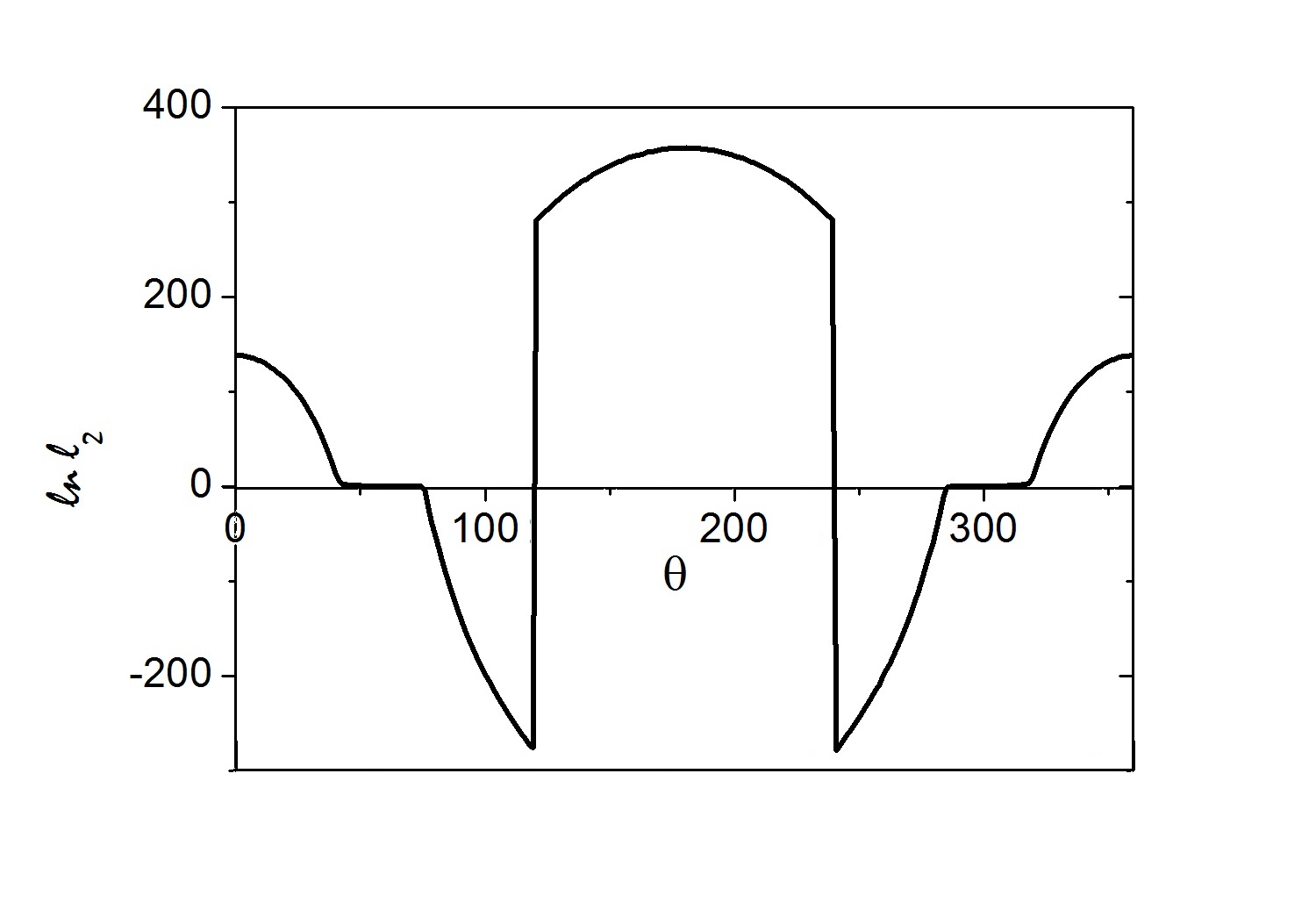}
             \includegraphics[width=5.5cm]{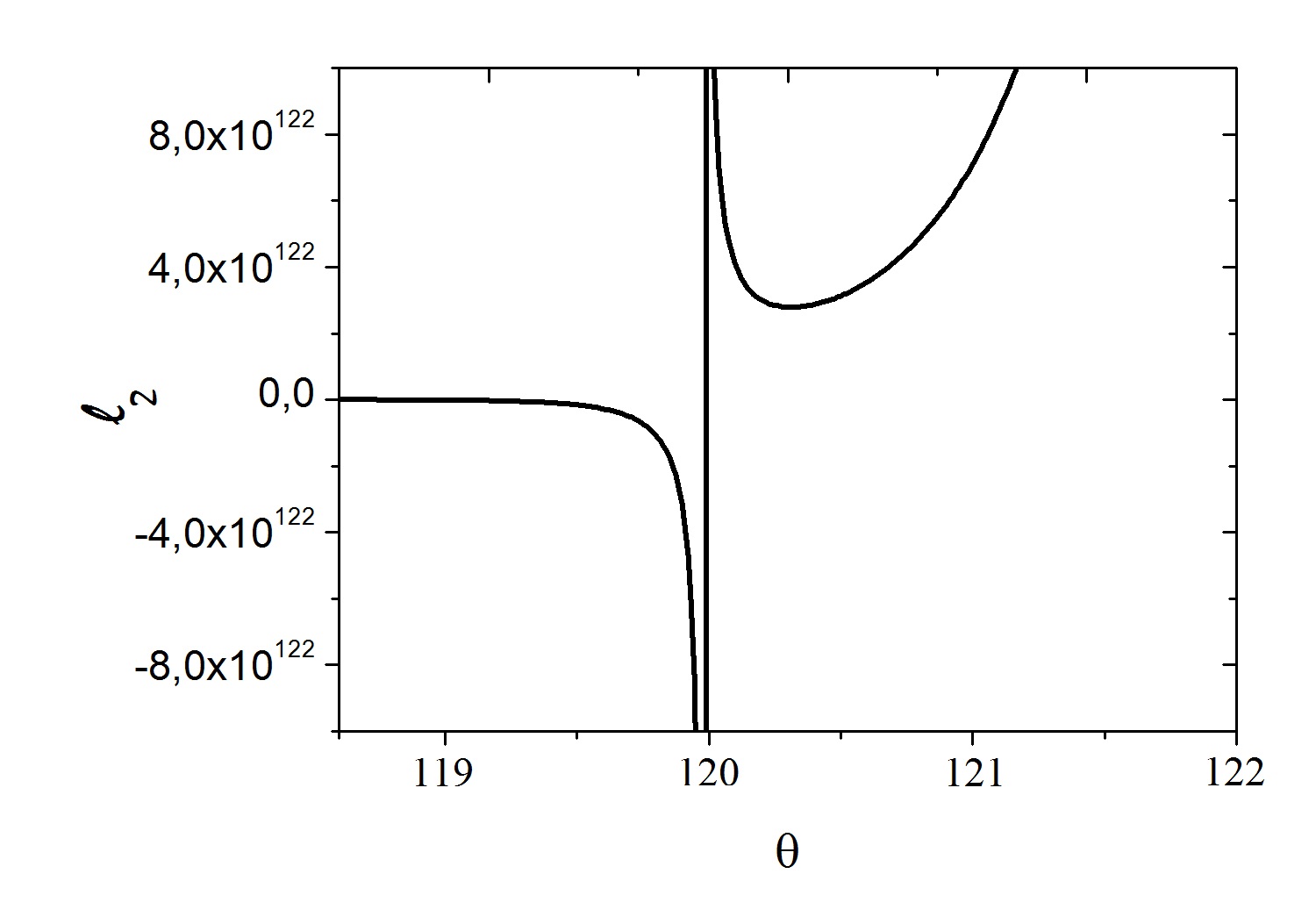}
             \caption{Dependence of $l_2$ parameter at  logarithmic (left)  and normal (right) scale. On the right side, we see the behavior of $l_2$ close to the point $\theta=\frac{2}{3}\pi$, where the denominator in (\ref{s22}) is zero.}\label{figL2}
           \end{figure*}

           \begin{figure*}[thbp]
             \includegraphics[width=5.5cm]{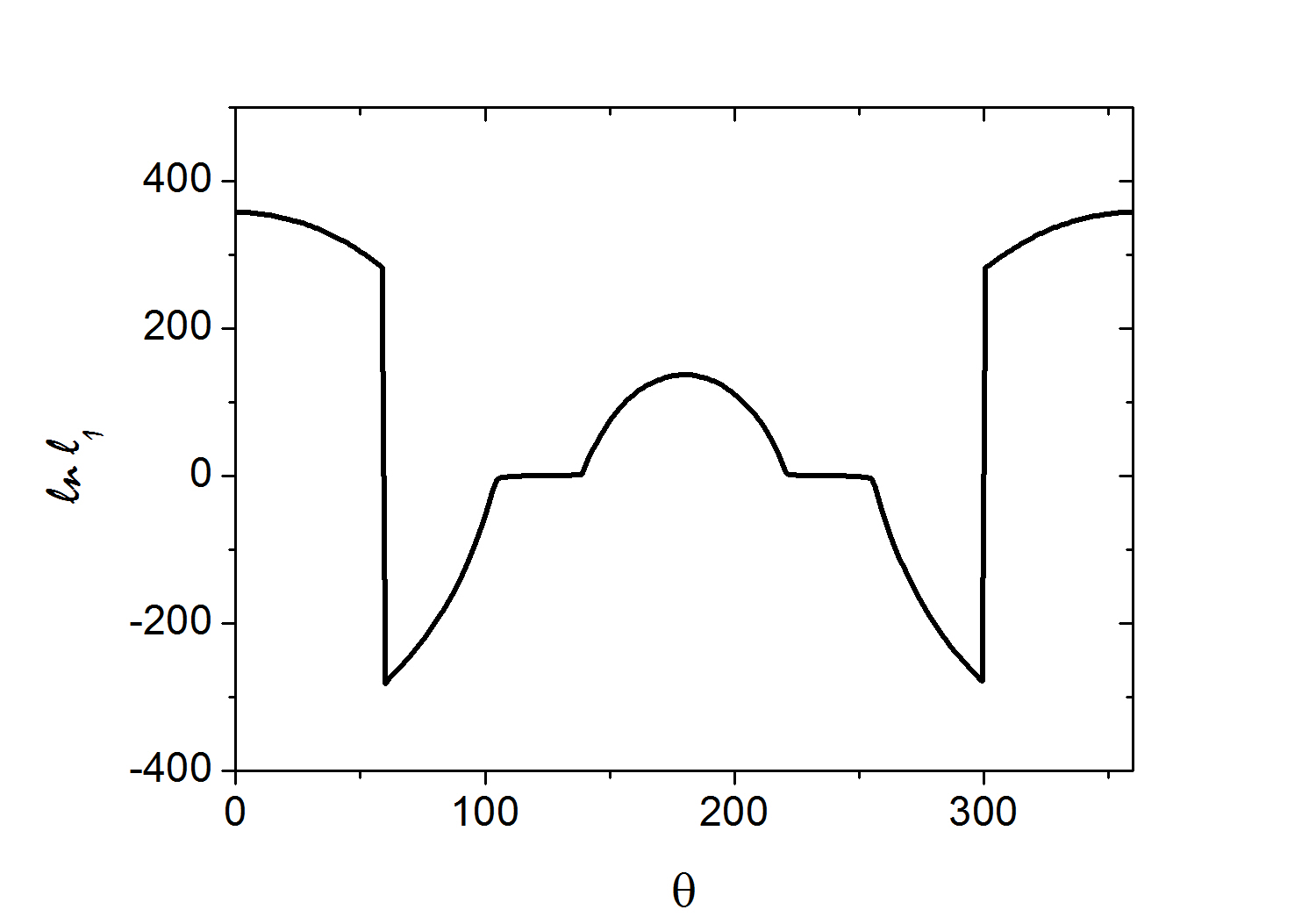}
             \includegraphics[width=5.5cm]{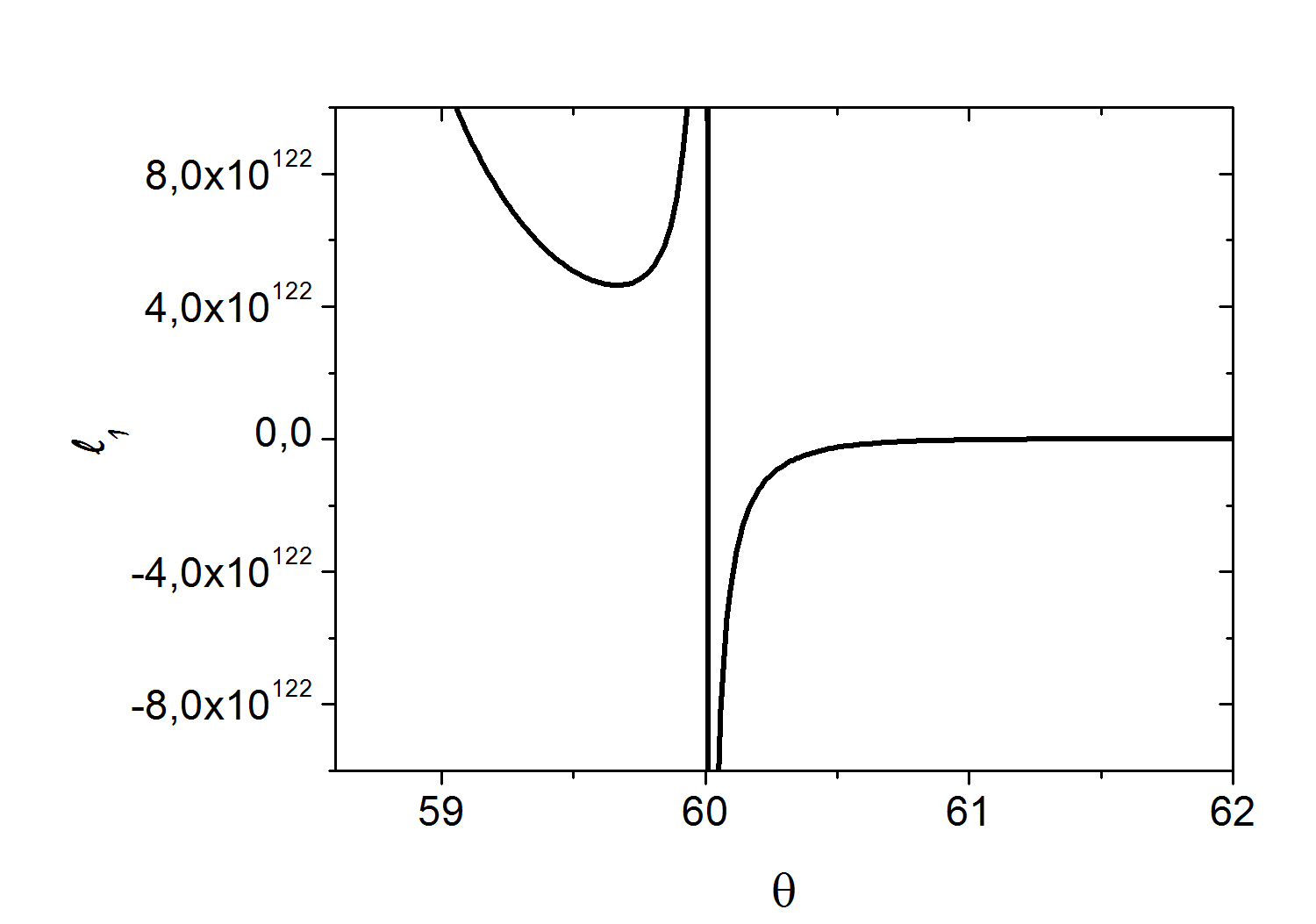}
             \caption{Dependence of $l_1$ parameter at  logarithmic (left)  and normal (right) scale. On the right side, we see the behavior of $l_1$ close to the point $\theta=\frac{1}{3}\pi$, where the denominator in (\ref{s22}) is zero.}\label{figL1}
           \end{figure*}

            where $\lambda\approx \frac{1}{m_{g}}$  and $m_{g}$ is the mass of gauge field. For $m_g=0$, the dependence of $l_2$ on $\theta$ we can see in Fig. \ref{figL2}. For better view on this dependence, it is depicted at the logarithmic scale as well (here we used the natural logarithm). Furthermore, there are 2 critical points, $\theta=2\pi/3$ and $\theta=4\pi/3$, for which the behavior of $l_2$ is not enough visible in this plot and that is why the graph on the right side is added. We see there that for $\theta=2\pi/3$ (as well as for $\theta=4\pi/3$), $l_2$ is not defined (which follows from the form of the denominator in (\ref{s22})) and no limit exists. For $\theta>2\pi/3$, $l_2$ strongly decreases from infinity to a local minimum at about $\theta=2.1\pi/3$ and then it increases up to the local maximum at $\theta=\pi$. In a similar way, the corresponding plots for $l_1$ are given in Fig. \ref{figL1}. The difference in comparison with the previous figure is the $60$ grades shift in the $\theta$ values.

             The solutions in (\ref{s22}) show that for small $m_{g}$, if the angle between sides of a defect be $\frac{\pi}{3}$, the couplings of anti-symmetric spins becomes more respect to couplings of parallel spins and the curvature of system becomes positive. In these conditions, electrons move in one special direction and conductivity emerges. This result is in agreement with previous predictions that for defect at an angle of $\frac{\pi}{3}$ between sides, the conductivity of system increases.


\section{anti-$F(R)$ gravity in a graphene }\label{o2}
Until now, we have known that for a defect with the angle near $\frac{\pi}{3}$, coupling of anti-parallel fermions tends to infinity and superconductivity emerges. Also,  for a defect with the angle near $\frac{2\pi}{3}$, coupling of parallel fermions tends to infinity, the quantity under $\sqrt{}$ in action (\ref{s19}), becomes negative which means that square energy of system becomes negative and some tachyonic states are produced which is the same of result in \cite{q6,q7,qq8}. We have:

  \begin{eqnarray}
  &&  \theta\rightarrow \frac{2\pi}{3}\Rightarrow l_{2}^{2},(l'_{2})^{2}\rightarrow \infty\Rightarrow\nonumber\\&&
 ( 2l_{1}^{2}-2l_{2}^{2}+2(l'_{1})^{2}-2(l'_{2})^{2}-(l'_{1})^{2}l_{2}-(l'_{2})^{2}l_{1})\ll 0\nonumber\\&& \text{and }\lambda\approx \frac{1}{m_{g}},m_{g}\ll 1\Rightarrow \sqrt{-(...)}\label{s23}
   \end{eqnarray}

To remove these tachyonic states, we use of the method in \cite{q6,q7,qq8,q8} and show that gravity changes to anti-gravity which leads to conductivity in opposite direction. We assume that first electrons move from atom A to atom B and then by creation of tachyonic states they return and path is similar to compactified circle.  We define
${\displaystyle <\psi^{3}_{L}>=\frac{r}{l_{s}^{3/2}}}$ where $l_{s}$ is
the separation distance between atoms, r is distance from the center of hexagonal molecule. We show that upper or low spins will be omitted during this compacting. We can obtain \cite{q6,q7,qq8,q8} :

\begin{eqnarray}
&& [X^{a},X^{b},X^{c}]=F^{abc} \quad X^{b}X^{c}=\psi_{L}^{b}\psi_{U}^{c} \quad [X^{a},\psi_{L}^{b}]=\partial_{a,U}\psi_{L}^{b}
 \nonumber \\ \nonumber \\
&& \nonumber \\
\Sigma_{a,b,c=0}^{3} \langle
F^{abc},F_{abc}\rangle_{U}&=&\Sigma_{a,b,c=0}^{3}
\langle[X^{a},X^{b},X^{c}],[X_{a},X_{b},X_{c}]\rangle_{U}\nonumber \\
&=&
- \Sigma_{a,b,c,a'b'c'=0}^{3}\varepsilon_{abcD}\varepsilon_{a'b'c'G}^{D}X^{a}\psi_{L}^{b}\psi_{U}^{c}X_{a'}\psi_{L}^{b'}\psi_{U}^{c'} \nonumber \\
&=& -
6\Sigma_{a,b,a',b'=0}^{3}\varepsilon_{ab3D}\varepsilon_{a'b'3}^{D}X^{a}\psi_{L}^{b}\psi_{U}^{3}X_{a'}\psi_{L}^{b'}\psi_{U}^{3}
\nonumber \\
&-&  6\Sigma_{a,b,c,a',b',c'=0,\neq
3}^{2}\varepsilon_{abcD}\varepsilon_{a'b'c'}^{D}X^{a}\psi_{L}^{b}\psi_{U}^{c}X_{a'}\psi_{L}^{b'}\psi_{U}^{c'}
\nonumber \\&=&
 - 6\left(\frac{r^{2}}{l_{s}^{3}}\right)\Sigma_{a,b,a',b'=0}^{3}\varepsilon_{ab3D}\varepsilon_{a'b'3}^{D}X^{a}\psi_{L}^{b}X_{a'}\psi_{L}^{b'} \nonumber \\
&-&
6\Sigma_{a,b,c,a',b',c'=0,\neq3}^{2}\varepsilon_{abcD}\varepsilon_{a'b'c'}^{D}X^{a}X^{b}\psi_{L}^{b}\psi_{L}^{b'}  \nonumber \\
&=& -6\left(\frac{r^{2}}{l_{s}^{3}}\right)\Sigma_{a,b=0}^{2}[X^{a},\psi_{L}^{b}]^{2} + E_{Extra} \nonumber \\
&=&
-6\left(\frac{r^{2}}{l_{s}^{3}}\right)\Sigma_{a,b=0}^{2}\partial_{U}^{a}\psi_{L,b}\partial_{a,U}\psi_{L}^{b}
+ E_{Extra}\nonumber \\
&&\nonumber \\
\Sigma_{a,b,c=0}^{3} \langle F^{abc},F_{abc}\rangle_{L}&=&\Sigma_{a,b,c=0}^{3}
\langle[X^{a},X^{b},X^{c}],[X_{a},X_{b},X_{c}]\rangle_{L}=-6\left(\frac{r^{2}}{l_{s}^{3}}\right)\Sigma_{a,b=0}^{2}\partial_{L}^{a}\psi_{U,b}\partial_{a,L}\psi_{U}^{b}
+ E_{Extra}\nonumber \\
&&\nonumber \\
&&\nonumber \\
\partial_{a}\partial_{b}X^{i}\partial^{a}\partial^{b}X^{i}_{L}&=&\Sigma_{a,b,c=0}^{3}
\langle[X^{a},X^{b},X^{i}],[X_{a},X_{b},X_{i}]=-6\left(\frac{r^{2}}{l_{s}^{3}}\right)\Sigma_{a,b=0}^{2}[X^{a},\Psi_{L}^{i}]^{2} + E_{Extra}\nonumber \\
&=& -6\left(\frac{r^{2}}{l_{s}^{3}}\right)\Sigma_{a,b=0}^{2}\partial_{U}^{a}\Psi_{L,b}\partial_{a,U}\Psi_{L}^{b}
+ E_{Extra}\nonumber \\
&&\nonumber \\
&&\nonumber \\
\partial_{a}\partial_{b}X^{i}\partial^{a}\partial^{b}X^{i}_{U}&=&\Sigma_{a,b,c=0}^{3}
\langle[X^{a},X^{b},X^{i}],[X_{a},X_{b},X_{i}]=-6\left(\frac{r^{2}}{l_{s}^{3}}\right)\Sigma_{a,b=0}^{2}[X^{a},\Psi_{U}^{i}]^{2} + E_{Extra}\nonumber \\
&=& -6\left(\frac{r^{2}}{l_{s}^{3}}\right)\Sigma_{a,b=0}^{2}\partial_{L}^{a}\Psi_{U,b}\partial_{a,L}\Psi_{U}^{b}
+ E_{Extra}\label{s24}
\end{eqnarray}

This equation shows that by the emergence of anti-gravity, coupling of anti-spins are removed and only parallel spins remain. Also, the sign of these couplings reverses. In fact, the symmetry of system is broken for third time and terms that are related to absorption are removed. This means that the anti-gravity between parallel spins is the main reason for the emergence of Pauli exclusion principle. In this principle, parallel spins repel each other as due to anti-gravity between parallel spins.  Now, we can calculate the curvatures in this system:

\begin{eqnarray}
 && R^{\rho}_{\sigma\mu\nu}=\langle
 F^{\rho}\smallskip_{\sigma\lambda},F^{\lambda}\smallskip_{\mu\nu}\rangle=
 -6\left(\frac{r^{2}}{l_{s}^{3}}\right)\Sigma_{\rho,\mu=0}^{2}\partial_{U}^{\rho}\psi_{L,\sigma}\partial_{U,\mu}\psi_{L,\nu}-
 6\left(\frac{r^{2}}{l_{s}^{3}}\right)\Sigma_{\rho,\mu=0}^{2}\partial_{L}^{\rho}\Psi_{U,\sigma}\partial_{L,\mu}\Psi_{U,\nu}\nonumber\\&& \Rightarrow  R^{\rho}_{\sigma\mu\nu,anti-gravity}=-R^{\rho}_{\sigma\mu\nu,gravity} \nonumber\\&&\nonumber\\&& R_{MN}=R^{\rho}_{\rho\mu\nu}=R_{aa'}+R_{ia'}+R_{ij'}=R_{Free-Free}+R_{Free-Bound}+R_{Bound-Bound}\label{s25}
 \end{eqnarray}

 and
 \begin{eqnarray}
 &&\kappa^{\mu}_{\nu}=\delta^{\mu}_{\nu}-\sqrt{\delta^{\mu}_{\nu}-H^{\mu}_{\nu}}\nonumber\\
 && H= g^{\mu\nu}H_{\mu\nu}=
 \delta^{\mu}_{\nu}- 6\left(\frac{r^{2}}{l_{s}^{3}}\right)\Sigma_{a,b=0}^{2}\partial_{L}^{a}\Psi_{U,b}\partial_{a,L}\Psi_{R}^{b}-
 6\left(\frac{r^{2}}{l_{s}^{3}}\right)\Sigma_{\rho,\mu=0}^{2}\partial_{L}^{\rho}\Psi_{U,\sigma}\partial_{L,\mu}\Psi_{U,\nu}+... =\nonumber\\&& -(R_{Free-Free}+R_{Free-Bound}+R_{Bound-Bound})\nonumber\\&&\nonumber\\&&\nonumber\\&&\delta^{\rho\sigma}_{\mu\nu}\kappa^{\mu}_{\rho}\kappa^{\nu}_{\sigma}= (R_{Free-Free}+R_{Free-Bound}+R_{Bound-Bound})_{anti-gravity}=-R_{gravity}\label{s26}
  \end{eqnarray}

These results are different from results in previous section. This is because that for a defect with angle near $(\frac{2\pi}{3})$, parallel spins become close to each other and repel each other. In these conditions, gravity changes to anti-gravity ($R\rightarrow -R$) and couplings of anti-parallel spins are removed. Free electrons that move in one special direction, can't continue their paths and return. This is because that curvature has a direct relation with tensor of energy-momentum and by reversing the sign of gravity, the sign of applied momentum on electrons reverses and they return. To obtain the explicit form of $F(R)$-gravity, we use of the method in previous section and  obtain the relevant terms of determinant
 in action (\ref{ss5}) separately. Substituting equations (\ref{s24},\ref{s25} and \ref{s26})
 in determinants (\ref{s11}), we get:

 \begin{eqnarray}
 \det(F)=\delta_{\rho\sigma}^{\mu\nu}\langle
 F^{\rho\sigma}\smallskip_{\lambda},F^{\lambda}\smallskip_{\mu\nu}\rangle
 =-6\left(\frac{r^{2}}{l_{s}^{3}}\right)\delta_{\rho\sigma}^{\mu\nu}(R^{\rho\sigma}_{Free-Free,\mu\nu}+R^{\rho\sigma}_{Bound-Bound,\mu\nu}+R^{\rho\sigma}_{Free-Bound,\mu\nu})\label{s27}\end{eqnarray}

 \begin{eqnarray}
 && \det(P_{abc}[ E_{mnl}
 +E_{mij}(Q^{-1}-\delta)^{ijk}E_{kln}])=\nonumber\\&&\delta_{\rho\sigma}^{\mu\nu}
 \Biggl[(g^{\mu}_{\rho}g^{\nu}_{\sigma}+ \langle \partial^{b}\partial^{a}X^{i},\partial_{b}\partial_{a}X^{i}\rangle+..)-
 \nonumber\\&&\frac{(g^{\mu}_{\rho}g^{\nu}_{\sigma}+
 \langle \partial^{b}\partial^{a}X^{i},\partial_{b}\partial_{a}X^{i}\rangle+...)}{[(\lambda)^{2}
 \det([X^{j}_{\alpha}T^{\alpha},X^{k}_{\beta}T^{\beta},X^{k'}_{\gamma}T^{\gamma}])]}\Biggr]=\nonumber\\&&\delta_{\rho\sigma}^{\mu\nu}
  \Biggl[(g^{\mu}_{\rho}g^{\nu}_{\sigma}- 6\left(\frac{r^{2}}{l_{s}^{3}}\right)\Sigma_{a,b=0}^{2}\partial_{L}^{a}\Psi_{U,b}\partial_{a,L}\Psi_{R}^{b}-
   6\left(\frac{r^{2}}{l_{s}^{3}}\right)\Sigma_{\rho,\mu=0}^{2}\partial_{L}^{\rho}\Psi_{U,\sigma}\partial_{L,\mu}\Psi_{U,\nu}+..)-
  \nonumber\\&&\frac{(g^{\mu}_{\rho}g^{\nu}_{\sigma}
  -6\left(\frac{r^{2}}{l_{s}^{3}}\right)\Sigma_{a,b=0}^{2}\partial_{L}^{a}\Psi_{U,b}\partial_{a,L}\Psi_{R}^{b}-
   6\left(\frac{r^{2}}{l_{s}^{3}}\right)\Sigma_{\rho,\mu=0}^{2}\partial_{L}^{\rho}\Psi_{U,\sigma}\partial_{L,\mu}\Psi_{U,\nu}+...)}{[(\lambda)^{2}
  \det([X^{j}_{\alpha}T^{\alpha},X^{k}_{\beta}T^{\beta},X^{k'}_{\gamma}T^{\gamma}])]}\Biggr]=
  \nonumber\\&&
 \nonumber\\&& - 6\left(\frac{r^{2}}{l_{s}^{3}}\right) [R_{Free-Free}+R_{Free-Bound}+R_{Bound-Bound}]\times\nonumber \\&&
 \left(-1+\frac{1}{\left[(\lambda)^{2}\det([X^{j}_{\alpha}T^{\alpha},X^{k}_{\beta}T^{\beta},X^{k'}_{\gamma}T^{\gamma}])\right]}\right)=
 \nonumber\\&&  -6\left(\frac{r^{2}}{l_{s}^{3}}\right)[R_{Free-Free}+R_{Free-Bound}+R_{Bound-Bound}](-1+\frac{1}{m_{g}^{2}})\nonumber\\&&\nonumber\\&&\nonumber\\&&\det(Q)\sim
  (i)^{2}(\lambda)^{2}\det([X^{j}_{\alpha}T^{\alpha},X^{k}_{\beta}T^{\beta},X^{k'}_{\gamma}T^{\gamma}])\det(E)\sim
  \nonumber\\&&
  -[(\lambda)^{2}\det([X^{j}_{\alpha}T^{\alpha},X^{k}_{\beta}T^{\beta},X^{k'}_{\gamma}T^{\gamma}])]\det(g)=-m_{g}^{2}\det(g)\label{s28}
 \end{eqnarray}

 By substituting equations (\ref{s25}) ,(\ref{s26}), (\ref{s27}) and
 (\ref{s28})  into the action
 (\ref{ss5}) and puting $6(\frac{r^{2}}{l_{s}^{3}})=1$, we get:

 \begin{eqnarray}
 && L_{b}^{a}=\delta_{b}^{a}\Bigl[\Big((1-m_{g}^{2})[R_{Free-Free}+R_{Free-Bound}+R_{Bound-Bound}]\nonumber\\&&+
 m_{g}^{2}\lambda^{2}\delta_{\rho\sigma}^{\mu\nu}(R^{\rho\sigma}_{Free-Free,\mu\nu}+R^{\rho\sigma}_{Bound-Bound,\mu\nu}+R^{\rho\sigma}_{Free-Bound,\mu\nu})\Big)\Bigr]
 \label{s29}
 \end{eqnarray}

  This Lagrangian shows that the for a defect with the  angle  near ($\frac{2\pi}{3}$), the sign of gravity and also the shape of it changes at each point and electrons move in opposite directions. Consequently, the usual conductivity disappears and one new conductivity in opposite direction emerges. Now, we can calculate the relevant action for conductivity in graphene by substituting equation (\ref{s29}) in equation (\ref{ss5}) :

 \begin{eqnarray}
 S_{co-Graphene} &=& -\int d^{4}x \sum_{n=1}^{Q}
 \delta^{a_{1},a_{2}...a_{n}}_{b_{1}b_{2}....b_{n}}L^{b_{1}}_{a_{1}}...L^{b_{p}}_{a_{p}}=\nonumber\\&&
 - \int d^{4}x \sum_{n=1}^{Q}
 \delta^{a_{1},a_{2}...a_{n}}_{b_{1}b_{2}....b_{n}}\nonumber\\
 &&\Big( \Bigl[\Big((1-m_{g}^{2})[R_{Free-Free}+R_{Free-Bound}+R_{Bound-Bound}]\nonumber\\&&+
  m_{g}^{2}\lambda^{2}\delta_{\rho_{1}\sigma_{1}}^{\mu_{1}\nu_{1}}(R^{\rho_{1}\sigma_{1}}_{Free-Free,\mu_{1}\nu_{1}}+R^{\rho_{1}\sigma_{1}}_{Bound-Bound,\mu_{1}\nu_{1}}+R^{\rho_{1}\sigma_{1}}_{Free-Bound,\mu_{1}\nu_{1}})\Big)\Bigr]
 ^{b_{1}}_{a_{1}}\times
 \nonumber\\&&.......\times\nonumber\\
 && \Bigl[\Big(1-m_{g}^{2})[R_{Free-Free}+R_{Free-Bound}+R_{Bound-Bound}]\nonumber\\&&+
  m_{g}^{2}\lambda^{2}\delta_{\rho_{n}\sigma_{n}}^{\mu_{n}\nu_{n}}(R^{\rho_{n}\sigma_{n}}_{Free-Free,\mu_{n}\nu_{n}}+R^{\rho_{n}\sigma_{n}}_{Bound-Bound,\mu_{n}\nu_{n}}+R^{\rho_{n}\sigma_{n}}_{Free-Bound,\mu_{n}\nu_{n}})\Big)\Bigr]
 ^{b_{n}}_{a_{n}}\Big)^{\frac{1}{2}}\nonumber\\&=&- \int d^{4}x
 \Biggl[\sqrt{-g}\Bigl(\sum_{n=1}^{Q}\Big(1-m_{g}^{2})[R_{Free-Free}+R_{Free-Bound}+R_{Bound-Bound}]^{n}+\nonumber\\&&
 \sum_{n=1}^{Q}m_{g}^{2n}\lambda^{2n}\delta^{\rho_{1}\sigma_{1}...\rho_{n}\sigma_{n}}_{\mu_{1}\nu_{1}...\mu_{n}\nu_{n}}
 (R^{\rho_{1}\sigma_{1}}_{Free-Free,\mu_{1}\nu_{1}}+R^{\rho_{1}\sigma_{1}}_{Bound-Bound,\mu_{1}\nu_{1}}+R^{\rho_{1}\sigma_{1}}_{Free-Bound,\mu_{1}\nu_{1}})...\times\nonumber\\&&(R^{\rho_{n}\sigma_{n}}_{Free-Free,\mu_{n}\nu_{n}}+R^{\rho_{n}\sigma_{n}}_{Bound-Bound,\mu_{n}\nu_{n}}+R^{\rho_{n}\sigma_{n}}_{Free-Bound,\mu_{n}\nu_{n}})
 \Bigr)\Biggr]\Big)^{\frac{1}{2}} \label{s30}
 \end{eqnarray}

  This equation shows that for a defect with the angle near ($\frac{2\pi}{3}$),  the shape of action changes and the conductivity appears in opposite direction. This is because that at this angle, parallel spins become close to each other and anti-gravity emerges. This new form of gravity has a direct relation with tensor of energy-momentum ($T_{\mu\nu}=R_{\mu\nu}-\frac{1}{2}R g_{\mu\nu}$) and  causes that one momentum produces in opposite direction and the path of free electrons reverses. Now, we can obtain the shape of anti-$F(R)$-gravity in this system:

  \begin{eqnarray}
  &&  \delta^{\rho_{1}\sigma_{1}...\rho_{n}\sigma_{n}}_{\mu_{1}\nu_{1}...\mu_{n}\nu_{n}}
  R_{\rho_{1}\sigma_{1}}^{\mu_{1}\nu_{1}}...R_{\rho_{n}\sigma_{n}}^{\mu_{n}\nu_{n}}\rightarrow
  R^{n}\nonumber\\&&\Rightarrow S_{co-Graphene}= -\int d^{4}x
  \Biggl[\sqrt{-g}\Big(\sum_{n=1}^{Q}\Big(\Big((1-m_{g}^{2})^{n}
   [R_{Free-Free}+R_{Free-Bound}+R_{Bound-Bound}]\nonumber\\&&+\lambda^{2n}m_{g}^{2n}[R_{Free-Free}+R_{Free-Bound}+R_{Bound-Bound}]^{n}\Big)\Biggr]\Big)^{\frac{1}{2}}\nonumber\\&&\Rightarrow
  S_{co-Graphene}= -\int d^{4}x
  \Biggl[\sqrt{-g}\Big(F(R)\Big)\Biggr]\nonumber\\&&
  F(R)=\sum_{n=1}^{Q}\Big([(1-m_{g}^{2})^{n}+\lambda^{2n}m_{g}^{2n}]
     [R_{Free-Free}+R_{Free-Bound}+R_{Bound-Bound}]^{n}\Big)\label{s31}
  \end{eqnarray}

\begin{figure*}[thbp]
             \includegraphics[width=5.0cm]{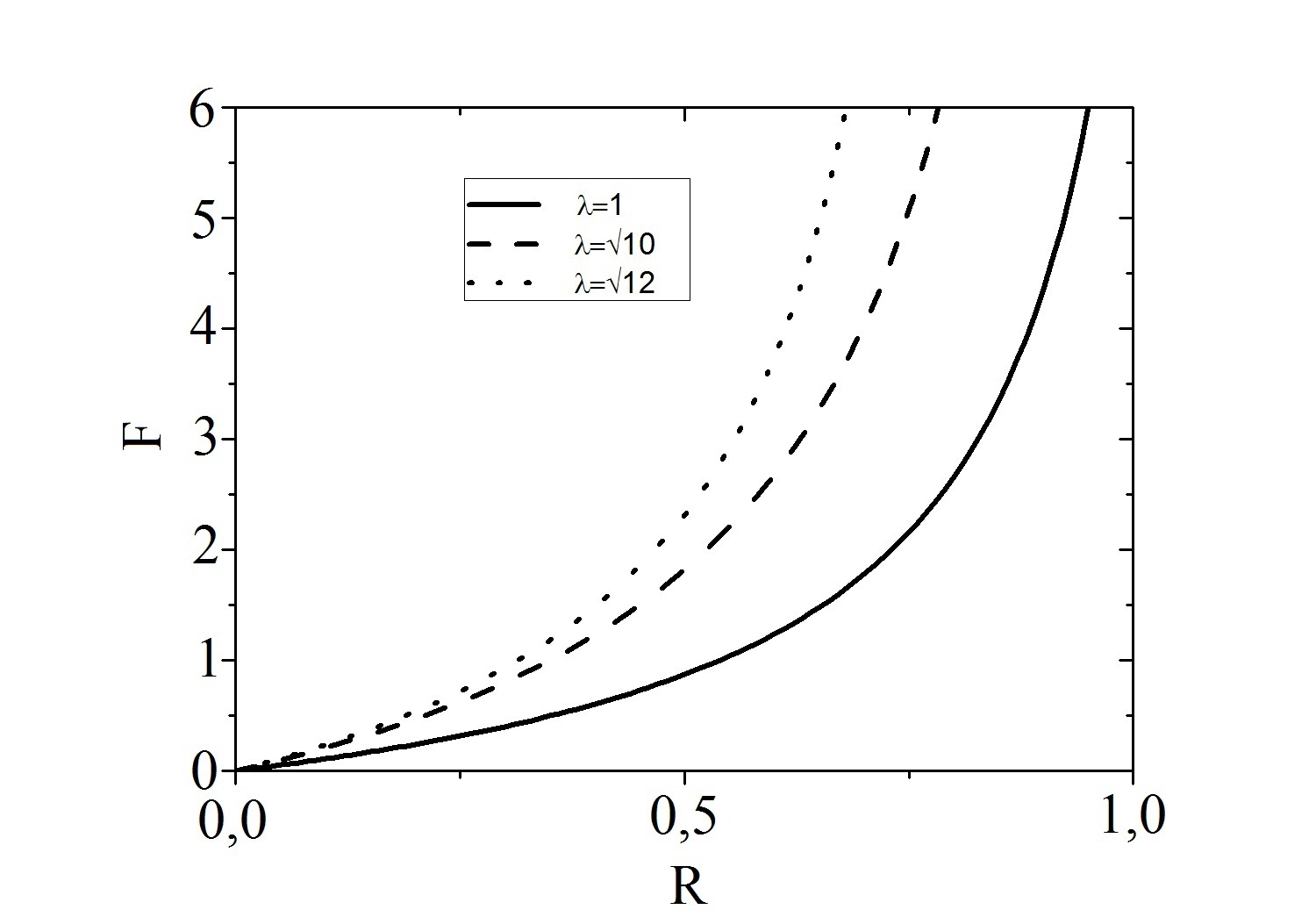}
              \includegraphics[width=5.0cm]{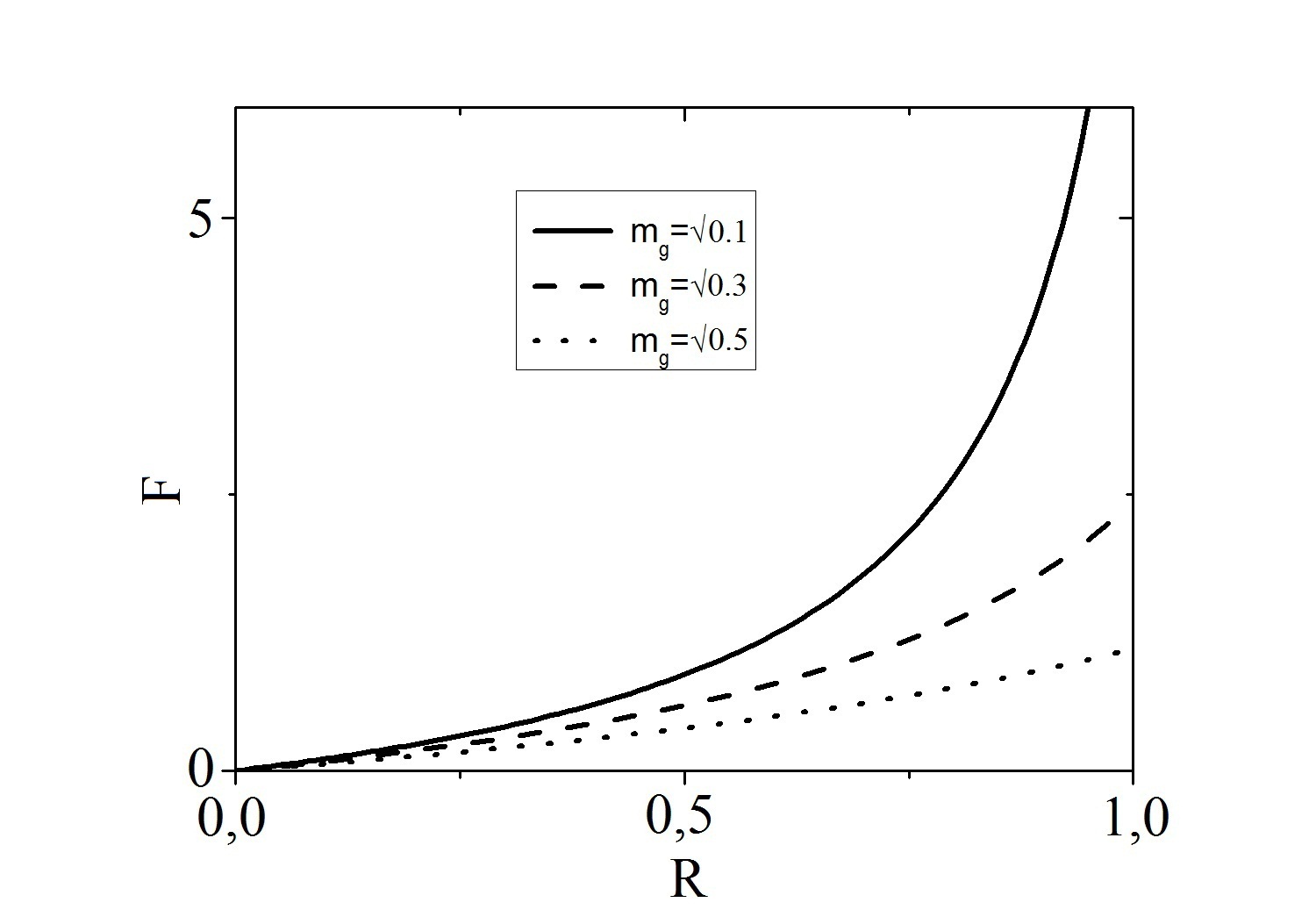}
               \includegraphics[width=5.0cm]{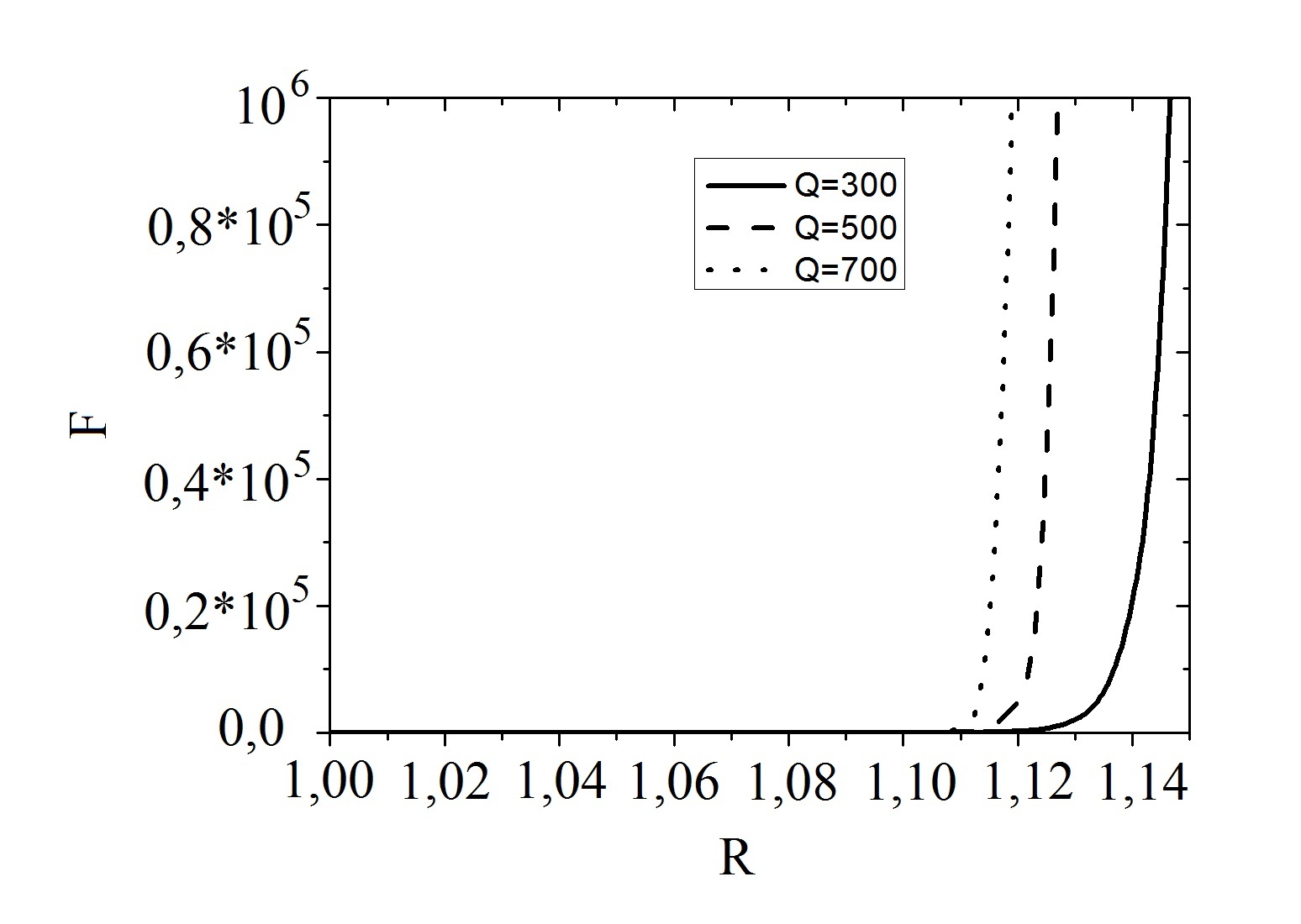}
             \caption{Dependence of antigravity in (35) on $R$ for different values of parameters $\lambda$ (left), $m_g$ (middle) and for different numbers of atoms in the structure (right).}\label{figSR}
           \end{figure*}

It is clear that action of (\ref{s31}) is very different of action (\ref{s18}) and derivatives of curvatures have been removed. This result is very the same of the result in \cite{q7,qq8} for branes. In M-theory and in these references \cite{q6,q7,qq8}, it has been shown  that when branes and anti-branes become very close to each other, the square energy of system becomes negative and some tachyonic states are appeared.  To remove these states, branes are compactified, gravity changes to anti-gravity and branes get away from each other. By compacting branes, the algebra of system changes and three dimensional brackets in Lie-three algebra are replace by two dimensional brackets in Lie-two algebra. Two dimensional brackets produce lower number of derivatives respect to three dimensional brackets. We will use of the same mechanism in graphene. When the angle between sheets of a defect becomes near $\frac{2\pi}{3}$, parallel spins are placed more closed to each other, respect to anti-parallel spins, gravity changes to anti-gravity and the algebra system transits from Lie-three algebra to Lie-two algebra with two dimensional brackets. In equation (\ref{s24}), it has been show that by reducing three dimensional brackets to two dimensional brackets, lower number of derivatives are appeared for spinors in expression of  ($\langle \partial^{b}\partial^{a}X^{i},\partial_{b}\partial_{a}X^{i}\rangle$) respect to (\ref{s9}) and in expression of ($\langle
  F^{\rho}\smallskip_{\sigma\lambda},F^{\lambda}\smallskip_{\mu\nu}\rangle$) respect to (\ref{s6} and\ref{s8}). Thus, by reduction of three dimensional brackets to two dimensional brackets, number of derivatives decreases and the shape of action changes. The dependence of the antigravity in this equation on $R$ we see in Fig. \ref{figSR}. This figure shows that in the critical points of the wormhole structure, the conductivity nearly does not depend on the number of the atoms in all the structure, the differences appear at greater curvatures.

As an example, we consider the emergence of $F(R)$-gravity in a graphene wormhole. If we suppose that the geometry of the wormhole can be described by the hyperboloid using the radius-vector
\begin{eqnarray}
\vec{r}=(r\cos\varphi,r\sin\varphi,\frac{c}{a}\sqrt{r^2-a^2}),
 \end{eqnarray}
where $c\ll a$, then the components of the metric tensor are
\begin{eqnarray}
g_{\varphi\varphi}=r^2,\hspace{1cm}g_{rr}=1+\frac{c^2}{a^2}\frac{r^2}{r^2-a^2},
 \end{eqnarray}
and the curvature $R=\frac{2c^2}{[(1+\frac{c^2}{a^2})r^2-a^2]^2}$. We see that for $c=0$, the parameters are changed into the planar case
\begin{eqnarray}
g_{\varphi\varphi}=r^2,\hspace{1cm}g_{rr}=1,\hspace{1cm}R=0.
 \end{eqnarray}
Using \cite{pert}, we can derive the relation between the parameters $c,a$ and the number of the defects in the wormhole center,
\begin{eqnarray}
\left(\frac{12}{N}\right)^2-\frac{c^2}{a^2}=1.
 \end{eqnarray}
Then, the curvature can be also expressed as
\begin{eqnarray}
R=\frac{2\left[(\frac{12}{N})^2-1\right]}{a^2[(\frac{12}{N})^2\frac{r^2}{a^2}-1]^2}.
\end{eqnarray}

\begin{figure*}[thbp]
             \includegraphics[width=5.0cm]{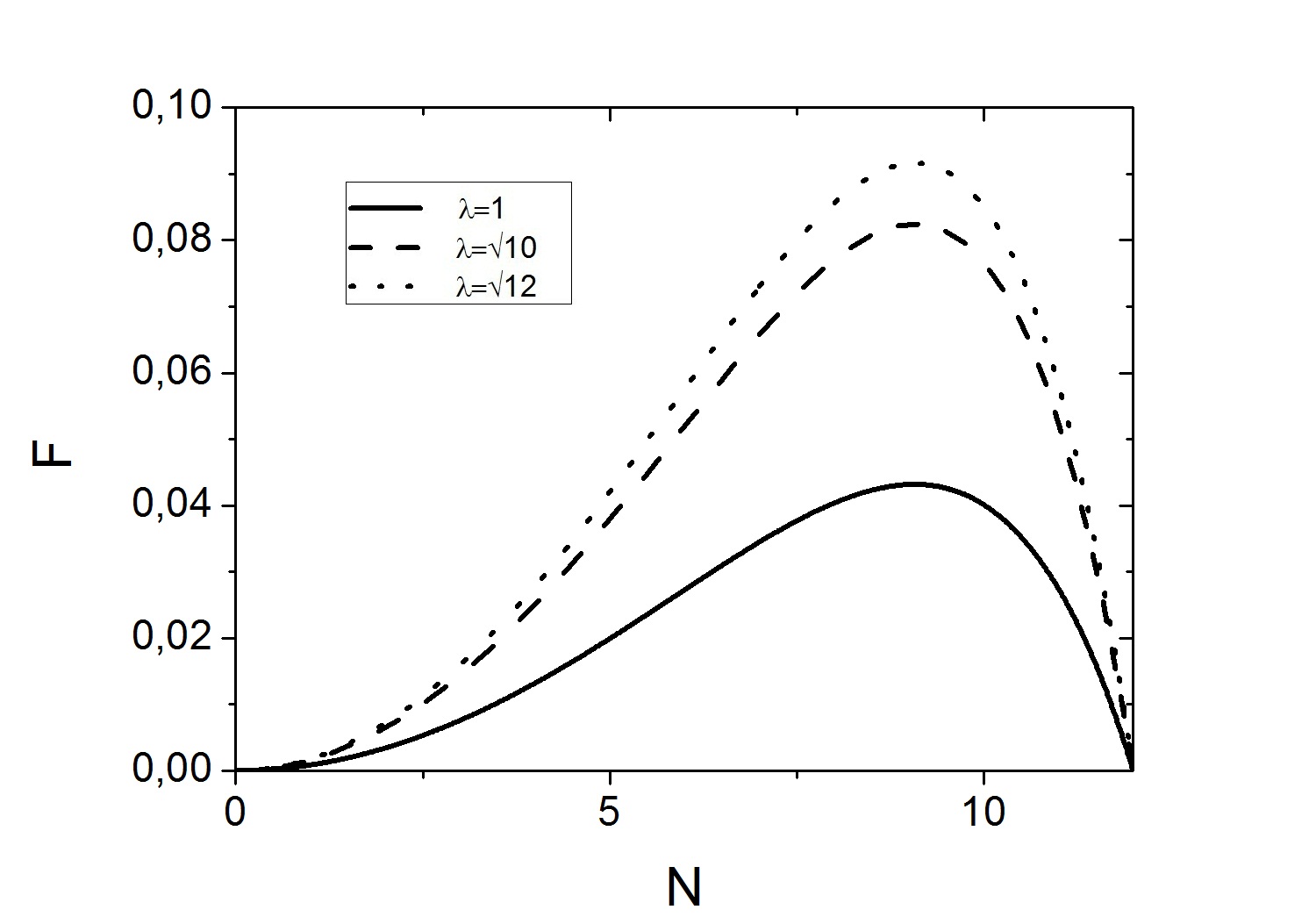}
              \includegraphics[width=5.0cm]{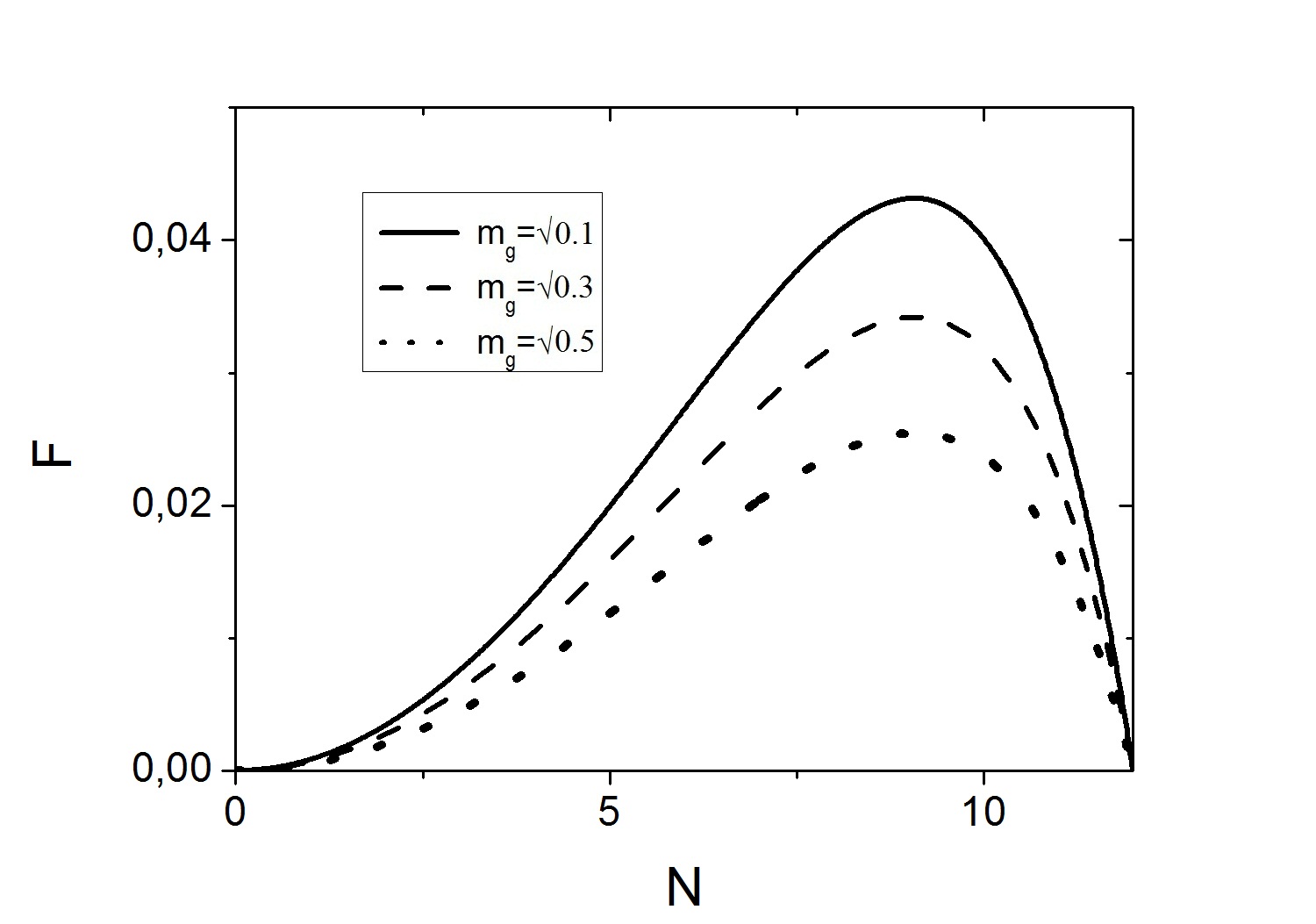}
               \includegraphics[width=5.0cm]{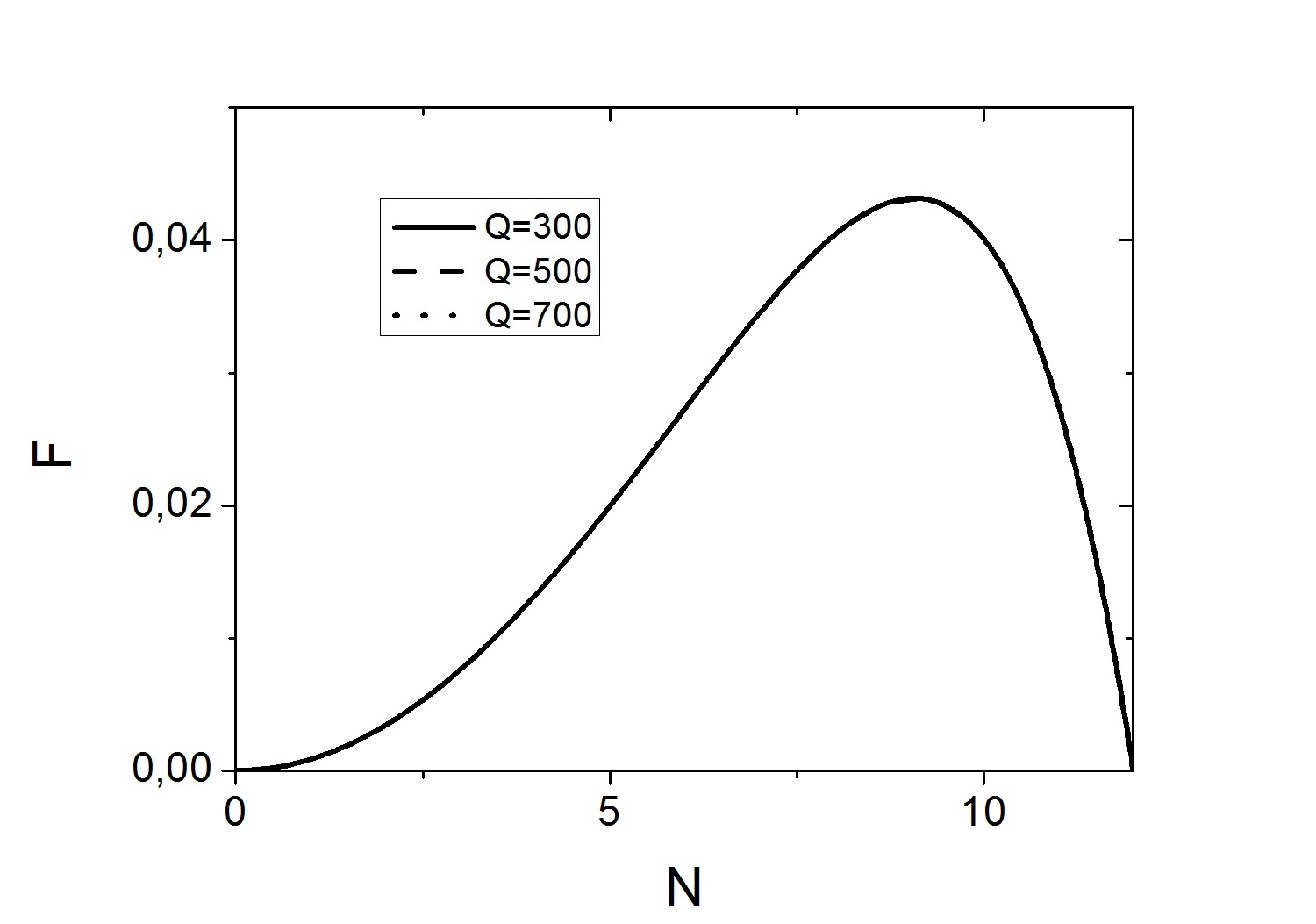}
             \caption{Dependence of antigravity in (35) on the number of the defects.}\label{figSR1}
           \end{figure*}

\begin{figure*}[thbp]
             \includegraphics[width=5.0cm]{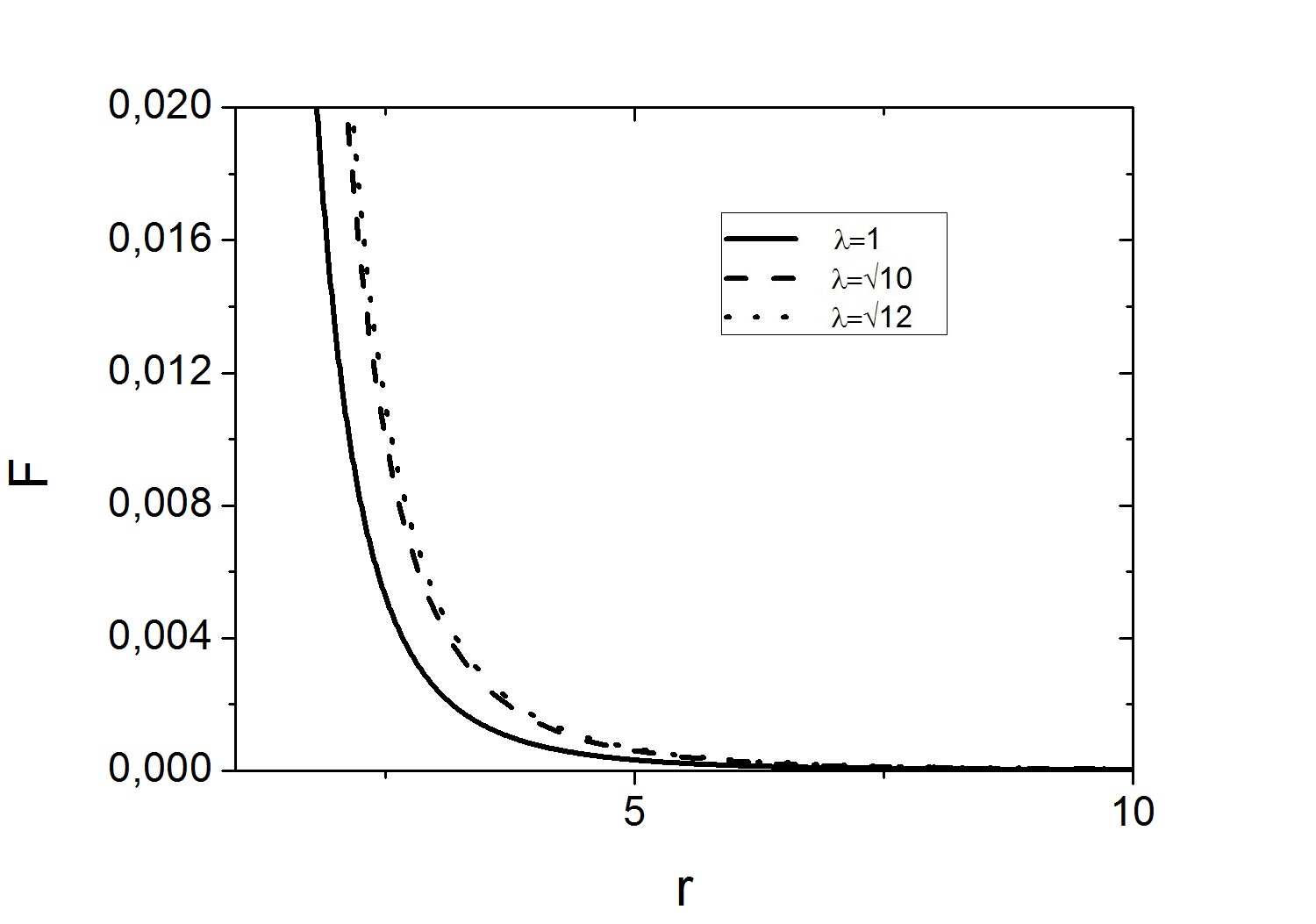}
              \includegraphics[width=5.0cm]{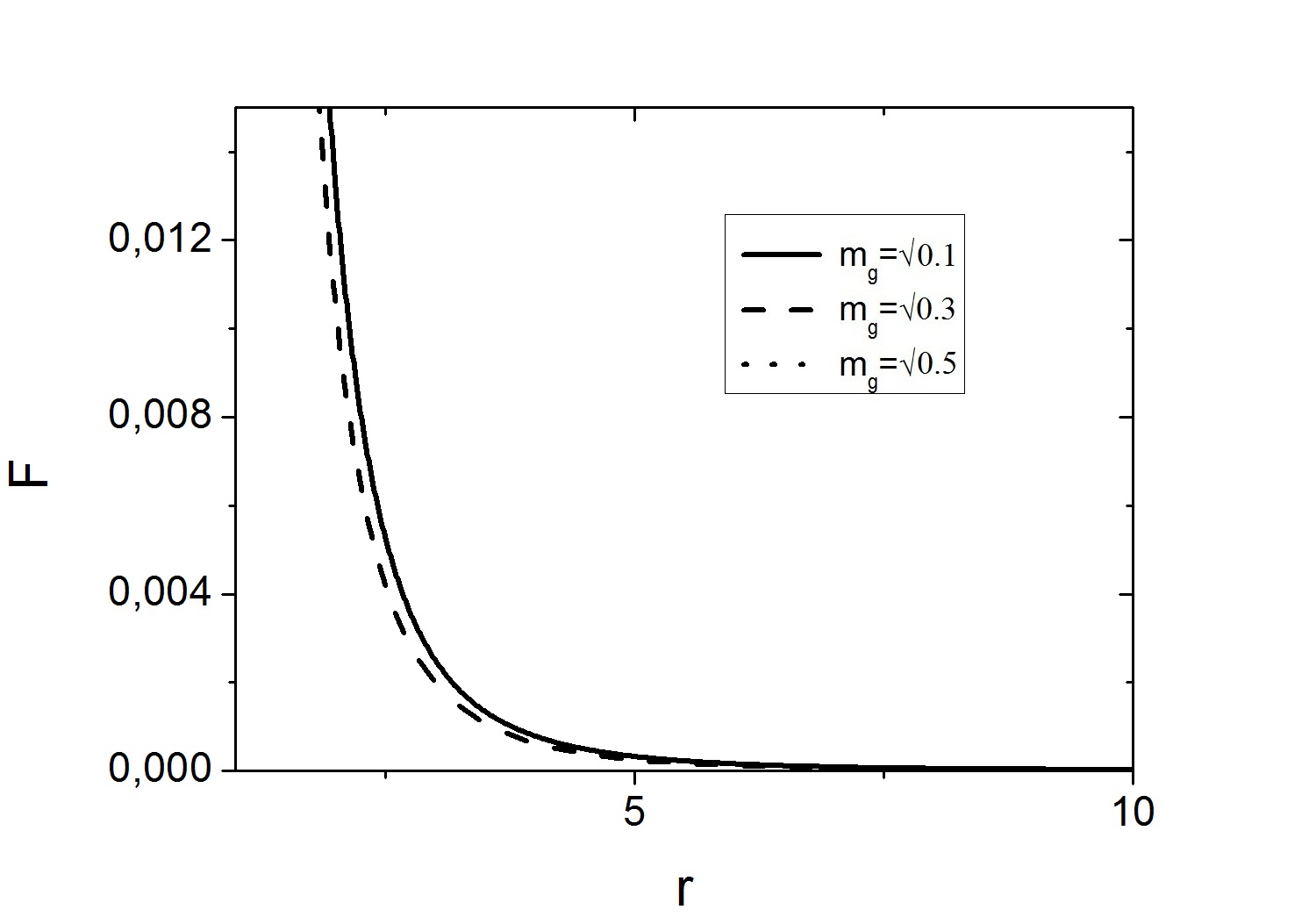}
               \includegraphics[width=5.0cm]{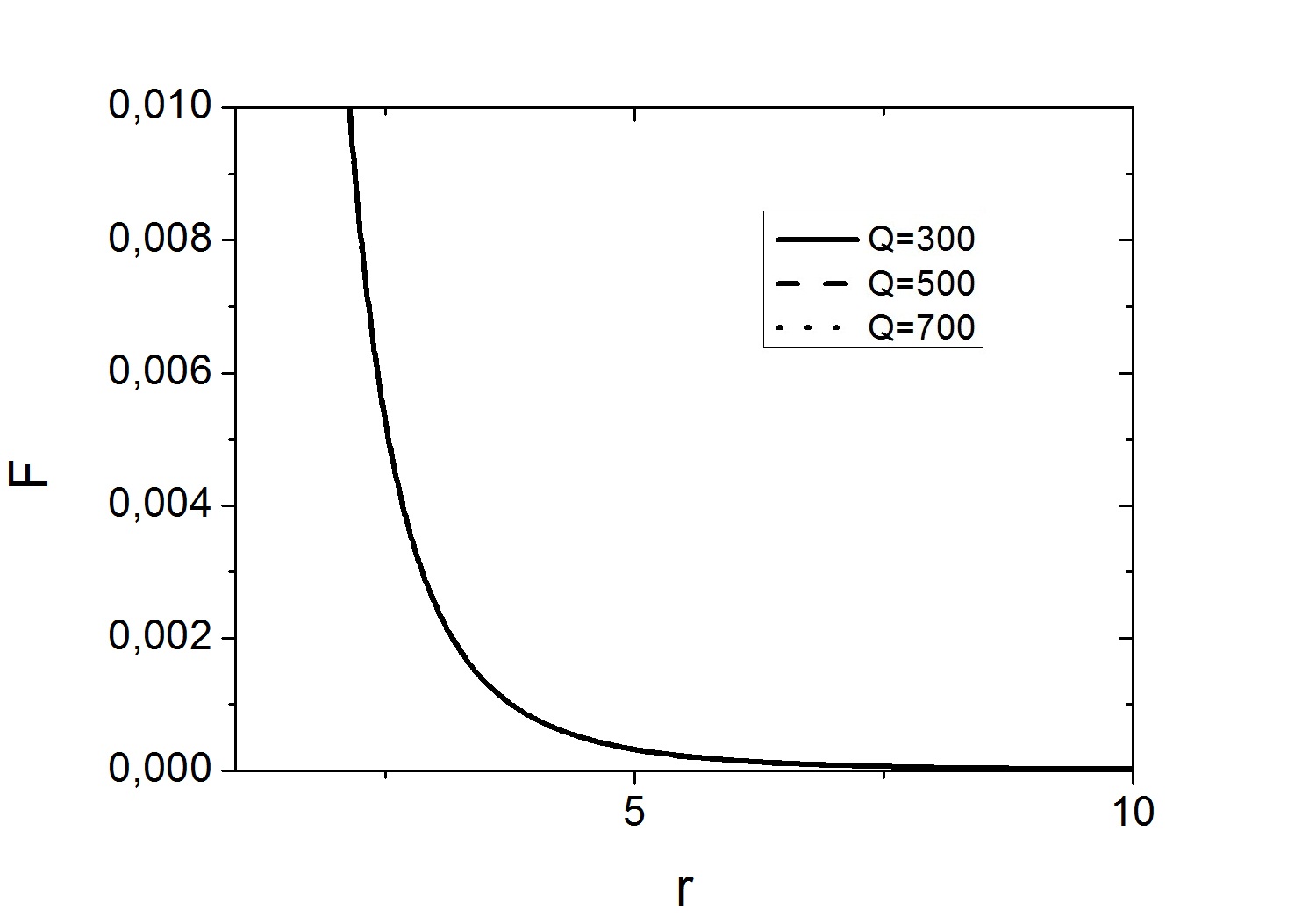}
             \caption{Dependence of antigravity in (35) on the distance from the wormhole center.}\label{figSR2}
           \end{figure*}

 In Figs. \ref{figSR1} and \ref{figSR2}, for the same values of the parameters, we see the dependence of the $F(R)$-gravity on the number of the defects and on the distance from the wormhole center. Similarly as in Fig. \ref{figSR}, the dependence on the number of the atoms in all the structure is negligible. Furthermore, Fig. \ref{figSR1} shows that the highest value of the $F(R)$-gravity could be in the case of 9 defects in the wormhole center. Also, Fig.\ref{figSR2} shows that by increasing distance from the wormhole center, the effect of defects decreases and $F(R)$-gravity decreases and shrinks to zero.

\begin{figure*}[thbp]
               \includegraphics[width=8cm]{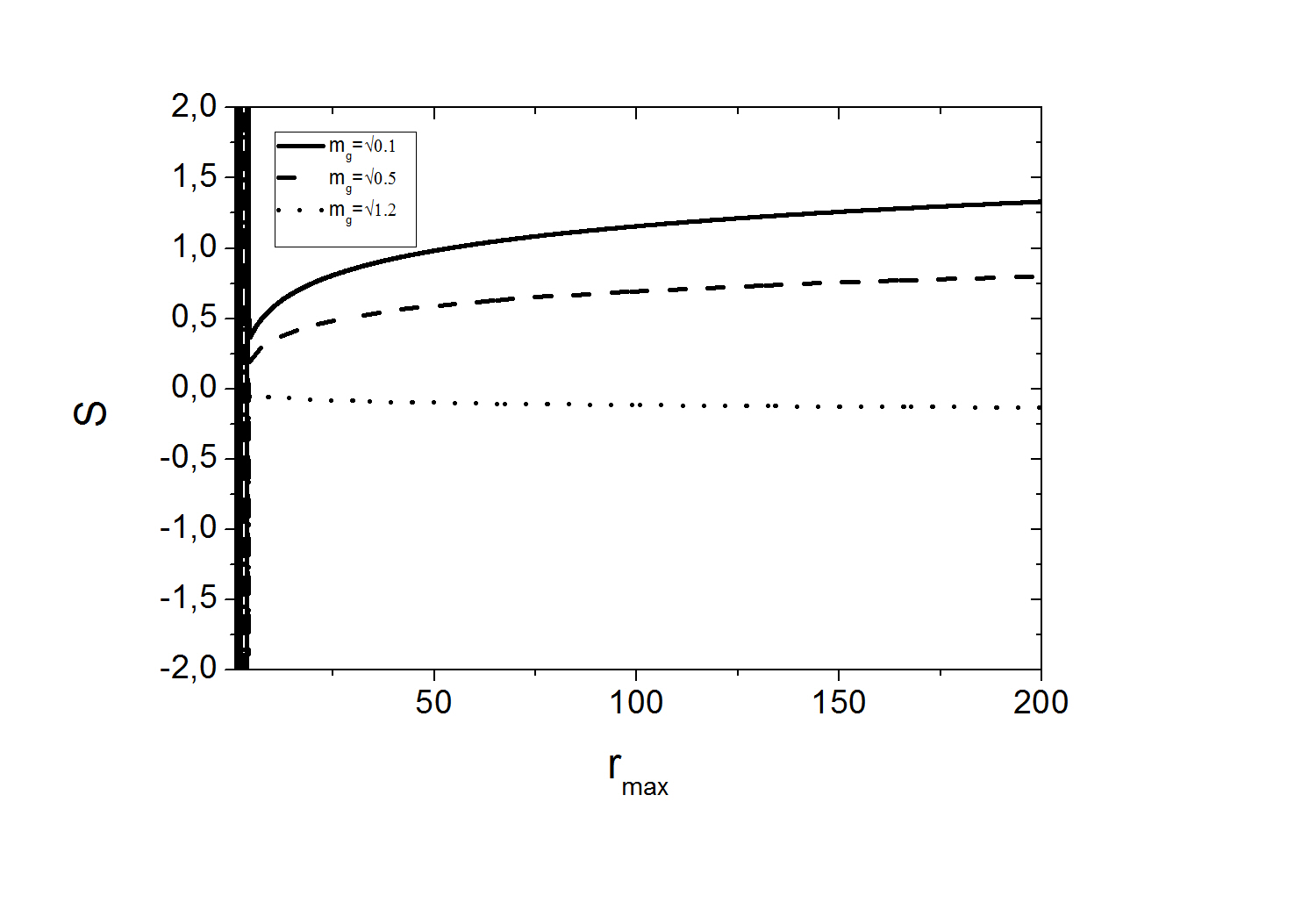}
             \caption{Dependence of conductivity in (35) on the size of the wormhole structure for different values of the parameter $m_g$.}\label{figSSR}
           \end{figure*}

 In Fig. \ref{figSSR}, we see the dependence of the of the conductivity on the size of the wormhole structure ($r_{max}=a$) for different values of the mass of the gauge field. It is clear that for bigger wormhole, the curvature of a graphene grows and  the conductivity increases.

The equation (35) also shows that if the angle between sides of a defect is near ($\frac{2\pi}{3}$), the shape of $F(R)$-gravity changes and includes lower orders of curvatures. This is because that in hexagonal molecule, anti-parallel spins are located at an angle ($\frac{\pi}{3}$) in neighbor atoms and  parallel spins are located at ($\frac{2\pi}{3}$). For ($\frac{\pi}{3}$), parallel spins become closed to each other and repel each other as due to anti-gravity between them.  Also, the effect of anti-parallel spins in  curvatures are canceled and the sign of curvature reverses which is a signature of anti-gravity between parallel spins. The motion of these electrons will be in opposite direction respect to initial path and consequently the path of conductivity reverses.

\section{Summary and Discussion} \label{sum}
In this research, we have proposed a new mechanism which considers the evolution of a graphene from a semi-conductor to a superconductor. To this aim, first, we design a symmetrical system that all anti-parallel electrons are pair. Then, we break this symmetry and produce free electrons that move along the sheets of graphene.  We show that curvature produced by parallel spins are canceled by curvature produced by anti-parallel spins and total curvature of system is zero.  The curvature has a direct relation with tensor of energy-momentum and thus no momentum is applied to electrons and they move randomly. We break this symmetry by entering defects to system and show that for special angles, anti-parallel spins become more closed to each other and their curvatures become more stronger than curvatures of parallel spins and an $F(R)$  gravity emerges. This gravity applies a momentum to electrons and lead to their motion in an special path and emergency of superconductivity. For some other angles between sides of the defects, parallel spins become closed to each other and repel each other. In these conditions, a new type of $F(R)$  gravity emerges which contains lower orders of curvatures. Also, the couplings of anti-parallel spins in  curvatures are removed and the sign of couplings for parallel spins in the curvature reverses which is a signature of anti-gravity between parallel spins. At this stage, electrons move in opposite direction respect to initial path and consequently the path of conductivity reverses.

The motion of the electrons is also influenced by other factors \cite{pert}: the differences in Fermi level at different distances from the center of a graphene wormhole, the electron mass acquisition due to relativistic effects connected with the effect of antigravity and high velocity of the electrons close to wormhole center and the properties of the connecting nanotube as an individual graphitic nonstructural. The first of the mentioned effects causes the creation of the electron flux which is directed to the wormhole center and as the result, the electric charge is accumulated in the wormhole center (the effect of the so-called "graphene black hole"). This property plays a key role in the consideration of the pillared graphene as a material for the storage of the fuels of future (hydrogen etc., \cite{pillared}).


\section*{Acknowledgments}
\noindent The work of Alireza Sepehri has been supported
financially by Research Institute for Astronomy and Astrophysics
of Maragha (RIAAM),Iran under research project No.1/4165-14.
The work was partly supported by VEGA Grant No. 2/0009/16.
 R. Pincak would like to thank the TH division in CERN for hospitality.
 All authors contributed equally to the paper.

 \end{document}